\documentclass[longauth]{aa} 

\usepackage{graphicx}
\usepackage{subfig}
\usepackage{placeins}
\usepackage{tabularx}
\usepackage{txfonts}
\usepackage{hyperref}
\hypersetup{breaklinks=true,colorlinks=true,urlcolor=blue, linkcolor=blue, citecolor=blue}

\begin{document}

   \title{Disk Evolution Study Through Imaging of Nearby Young Stars
(DESTINYS): The SPHERE view of the Orion star-forming region \thanks{Based on observations collected at the European Southern Observatory under ESO programme(s) 0104.C-0195(A), 1104.C-0415 and 108.2216.001}}


   \author{P.-G. Valeg\r{a}rd\inst{1}
          \and
          C. Ginski\inst{2}
          \and
          A. Derkink\inst{1}
          \and
          A. Garufi\inst{3}
          \and
          C. Dominik\inst{1}
          \and
          \'A. Ribas\inst{4}
          \and
          J. P. Williams\inst{5}
          ,
          M. Benisty\inst{6}
          ,
          T. Birnstiel\inst{7,8}
          ,
          S. Facchini\inst{9}
          ,
          G. Columba\inst{10,11}
          ,
          M. Hogerheijde\inst{1,12}
          ,
          R.\,G.~van~Holstein\inst{13}
          ,
          J. Huang\inst{14}
          ,
          M. Kenworthy\inst{12}
          ,
          C. F. Manara\inst{15}
          ,
          P. Pinilla\inst{17}
          ,
          Ch. Rab\inst{7,18}
          ,
          R. Sulaiman\inst{19}
          ,
          A. Zurlo\inst{20,21,22}}

    \institute{Anton Pannekoek Institute for Astronomy (API), University of Amsterdam, Science Park 904, 1098 XH Amsterdam, The Netherlands, \email{p.g.valegard@uva.nl}
   \and School of Natural Sciences, Center for Astronomy, University of Galway, Galway, H91 CF50, Ireland
   \and INAF, Osservatorio Astrofisico di Arcetri, Largo Enrico Fermi 5, I-50125 Firenze, Italy
   \and Institute of Astronomy, University of Cambridge, Madingley Road, Cambridge, CB3 0HA, UK
   \and Institute for Astronomy, University of Hawaii, Honolulu, HI 96822, USA
   \and University of Grenoble Alps, CNRS, IPAG, 38000 Grenoble, France
   \and University Observatory, Faculty of Physics, Ludwig-Maximilians-Universität München, Scheinerstr. 1, 81679 Munich, Germany
    \and Exzellenzcluster ORIGINS, Boltzmannstr. 2, D-85748 Garching, Germany   
   \and Dipartimento di Fisica, Universit\`{a} degli Studi di Milano, Via Celoria 16, I-20133 Milano, Italy
   \and Department of Physics and Astronomy "Galileo Galilei" - University of Padova, Vicolo dell’Osservatorio 3, 35122 Padova, Italy
   \and INAF – Osservatorio Astronomico di Padova, Vicolo dell’Osservatorio 5, 35122 Padova, Italy
   \and Leiden Observatory, Leiden University, Niels Bohrweg 2, NL-2333 CA Leiden, The Netherlands
   \and European Southern Observatory, Alonso de C\'{o}rdova 3107, Casilla 19001, Vitacura, Santiago, Chile
   \and Department of Astronomy, Columbia University, 538 W. 120th Street, Pupin Hall, New York, NY 10027
   \and European Southern Observatory, Karl-Schwarzschild-Strasse 2, 85748 Garching bei M\"{u}nchen, Germany
   \and Max-Planck-Institut f\"{u}r Astronomie, K\"{o}nigstuhl 17, 69117, Heidelberg, Germany
    \and Mullard Space Science Laboratory, University College London, Holmbury St Mary, Dorking, Surrey RH5 6NT, UK 
    \and Max-Planck-Institut für extraterrestrische Physik, Giessenbachstrasse 1, 85748 Garching, Germany
    \and Department of Physics, American University of Beirut, PO Box 11-0236, Riad El-Solh, Beirut 11097 2020, Lebanon  
    \and Instituto de Estudios Astrof\'isicos, Facultad de Ingenier\'ia y Ciencias, Universidad Diego Portales, Av. Ej\'ercito Libertador 441, Santiago, Chile
    \and Escuela de Ingenier\'ia Industrial, Facultad de Ingenier\'ia y Ciencias, Universidad Diego Portales, Av. Ej\'ercito Libertador 441, Santiago, Chile
    \and Millennium Nucleus on Young Exoplanets and their Moons (YEMS)
   }

\titlerunning{Infrared evolution study Orion}
\authorrunning{Valeg\r{a}rd et al.}             

   \date{January 2022}

 
  \abstract
   {Resolved observations at near-infrared (near-IR) and millimeter wavelengths have revealed a diverse population of planet-forming disks. In particular, near-IR scattered light observations usually target close-by, low-mass star-forming regions. However, disk evolution in high-mass star-forming regions is likely affected by the different environment. Orion is the closest high-mass star-forming region, enabling resolved observations to be undertaken in the near-IR.}
   {We seek to examine planet-forming disks, in scattered light, within the high-mass star-forming region of Orion in order to study the impact of the environment in a higher-mass star-forming region on disk evolution.}
   {We present SPHERE/IRDIS H-band data for a sample of 23 stars in the Orion star-forming region observed within the DESTINYS (Disk Evolution Study Through Imaging of Nearby Young Stars) program. We used polarization differential imaging in order to detect scattered light from circumstellar dust. From the scattered light observations we characterized the disk orientation, radius, and contrast. We analysed the disks in the context of the stellar parameters and the environment of the Orion star-forming region. We used ancillary X-shooter spectroscopic observations to characterize the central stars in the systems. We furthermore used a combination of new and archival ALMA mm-continuum photometry to characterize the dust masses present in the circumstellar disks.} 
   {Within our sample, we detect extended circumstellar disks in ten of 23 systems. 
   Of these, three are exceptionally extended (V351\,Ori, V599\,Ori, and V1012\,Ori) and show scattered light asymmetries that may indicate perturbations by embedded planets or (in the case of V599\,Ori) by an outer stellar companion.
   Our high-resolution imaging observations are also sensitive to close (sub)stellar companions and we detect nine such objects in our sample, of which six were previously unknown. We find in particular a possible substellar companion (either a very low-mass star or a high-mass brown dwarf) 137\,au from the star RY\,Ori. \\
   We find a strong anticorrelation between disk detection and multiplicity, with only two of our ten disk detections located in stellar multiple systems.
   We also find a correlation between scattered light contrast and the millimeter flux. This trend is not captured by previous studies of a more diversified sample and is due to the absence of extended, self-shadowed disks in our Orion sample. Conversely, we do not find significant correlations between the scattered light contrast of the disks and the stellar mass or age. We investigate the radial extent of the disks and compare this to the estimated far-ultraviolet (FUV) field strength at the system location. While we do not find a direct correlation, we notice that no extended disks are detected above an FUV field strength of $\sim 300$G$_0$. }
   {}

 
   \maketitle
%

\section{Introduction}
The structure and evolution of circumstellar disks around young pre-main-sequence
(PMS) stars holds important keys to the understanding of planet formation and the origin of the architecture of planetary systems. Over the last decade, spatially resolved imaging of disks around PMS stars, from (sub)millimeter wavelengths with ALMA to optical and near-infrared (near-IR) wavelengths with Subaru, GPI, and VLT/SPHERE, has shown that disks come in a variety of shapes and sizes. Substructures in the form of cavities, gaps, rings, arcs, and spiral arms are routinely observed \citep{2018ApJ...869L..41A, 2020A&A...633A..82G, 2020ARA&A..58..483A,2020IAUS..345...96P, 2023ASPC..534..605B}. 

The study of protoplanetary disks and disk evolution using spatially resolved observations is, because of the resolution limit of current instrumentation, mostly constrained to the solar neighbourhood. At larger distances, $\geq 500\ \rm pc$, only the largest disk features such as gaps and rings on scales of 10 au or larger can be resolved with the resolution currently available. Observations have therefore predominantly focused on the closest star-forming regions, where high contrast and resolution is easier to achieve. These nearby star-forming regions (Taurus, Auriga, Lupus, Chameleon, and Ophiuchus) are all low-mass star-forming regions (LM-SFRs) with relatively low stellar densities. Out to a distance of $500\ \rm pc$ from the Sun, the Orion star-forming region is the only high-mass star-forming region (HM-SFR.) The initial mass function tells us that in a HM-SFR we can find not only more stars, but also stars that start out with higher masses. Protoplanetary disk masses scale with stellar mass \citep{2013ApJ...771..129A, 2013ApJ...773..168M, 2016ApJ...831..125P, 2016ApJ...828...46A, 2023ASPC..534..539M}, and thus one can also expect that the most massive disks, at least initially, are more massive in a HM-SFR than those in a LM-SFR. This makes disk observations especially interesting to study, since Orion's higher stellar density provides a different environment for the evolution of protoplanetary disks \citep{2018MNRAS.478.2700W, 2022EPJP..137.1132W}.

The most luminous and massive stars in the Orion star-forming region locally dominate their environment, for example $\theta_{1}$\,Ori C (O6Vp+B0V), $\sigma$\,Ori (O9.5V+B0.5V), and NU Ori (O9.5), with their strong ultraviolet radiation fields (UV fields.) In the LM-SFR, the most massive stars are Herbig and intermediate-mass T Tauri stars that, in contrast to more luminous and more massive O stars, have less impact on their immediate surroundings due to a much weaker UV field. Observations show that the impact of external irradiation, and therefore external photo-evaporation, drops off quickly as a function of distance \citep{2014ApJ...784...82M, 2017AJ....153..240A, 2018MNRAS.478.2700W, 2022EPJP..137.1132W}. Proplyds in the environment close to $\theta_{1} Ori$, albeit small, still exist even though they are expected to dissipate quickly. Their existence can be explained by an extended period of star formation \citep{2019MNRAS.490.5478W}. Millimeter surveys of the $\sigma$\,Ori cluster show a trend that the smallest disks are also the disks closest to the OB system at its center ($\leq 0.5$\,pc) \citep{2017AJ....153..240A}. This is also true for the Orion Nebula Cluster (ONC) \citep{1987ApJ...321..516C, 1994ApJ...436..194O, 1998AJ....116..293B, 2008AJ....136.2136R, 2016ApJS..226....8K, 2018ApJ...860...77E} and for NGC 2244 \citep{2007ApJ...660.1532B,2008ApJ...688..408B}. At larger distances from the cluster center, $\geq 4\ \rm pc$, there seems to be little difference in the disk dust mass between HM-SFR and LM-SFR \citep{2011A&A...525A..81R, 2022A&A...661A..53V}; therefore, the initial disk mass, if no external irradiation is assumed, is expected to be similar to the currently measured one. In that case, stellar age has a larger impact on the disk dust mass than external irradiation \citep{2019A&A...628A..85V}. 


Another important factor for disk evolution is stellar multiplicity, since companions can truncate the young disk and induce changes to the disk structure such as spiral arms, warps, and misalignments in the disks. For very close binaries, the disk can form around both stars, a so-called circumbinary disk \citep{2019ApJ...872..158A, 2019A&A...628A..95M, 2022A&A...662A.121R, 2021MNRAS.501.2305Z}. Studies of stellar multiplicity as a function of mass suggest that multiplicity increases with the stellar mass of the primary component (see \citealt{2013ARA&A..51..269D} or \citealt{2023ASPC..534..275O}.) In the formation of close binaries, the binary frequency does not vary much in the mass range below $1.5\ \rm M_{\odot}$, while for intermediate- and high-mass stars there is a higher frequency \citep{2013ARA&A..51..269D}. While stellar mass companions and brown dwarfs might be easier to detect, substellar companions are scarcely detected in young systems using infrared (IR) observations \citep{2022NatAs...6..639G}. At these wavelengths it is very challenging to resolve embedded planets since the disk outshines the scattered light of the planet \citep{2018A&A...617A..44K, 2022NatAs...6..639G}. At this point in time there are only a couple of confirmed cases, HD 169142b \citep{2013ApJ...766L...2Q, 2014ApJ...792L..22B, 2014ApJ...792L..23R}, which recently has been re-detected \citep{2023MNRAS.522L..51H}, PDS 70b \citep{2018A&A...617A..44K}, and AB Aur b \citep{2022NatAs...6..751C}, which has come under discussion as a false positive \citep{2023AJ....166..220Z}. The confirmation of these three planets is in contention with the disk substructures and, as discussed with AB Aur b, only underlines the difficulty of observing embedded planets in protoplanetary disks. 


Because of the distance to the Orion star-forming region, there are only few spatially resolved observations of disks in scattered light in this region such as, for example, HD34282 \citep{2021A&A...649A..25D} and V1247 Ori \citep{2016PASJ...68...53O}. In this paper, we present near-IR observations of 23 stars in Orion with no prior SPHERE data. The majority of these were observed within the Disk Evolution Study Through Imaging of Nearby Young Stars (DESTINYS) large program \citep{2020A&A...642A.119G, 2021ApJ...908L..25G}, in which 85 targets from six different star-forming regions were probed to characterize the scattered light from the circumstellar material, and to constrain its evolution throughout the various stages of planet formation. Companion papers to this study on the low-mass Taurus and Chamaeleon I star-forming regions are presented in Garufi et al. (submitted) and Ginski et al. (submitted.)


The paper is structured in the following way. The observations and data reduction are described in Sect.\,\ref{sec:observations}, while the sample is presented in Sect.\,\ref{sec:sample}. Using the SPHERE images, we report on the detection of stellar companions and circumstellar disks as well as on the correlation between the disk brightness and any stellar and disk properties in Sect.\,\ref{sec:results}. We then discuss the implications of the data as a whole and conclude in Sects.\,\ref{sec:discussion} and \ref{sec:conclusions}, respectively.

\section{Observations and data reduction} \label{sec:observations}

\subsection{SPHERE polarimetric imaging}
The SPHERE \citep{2019A&A...631A.155B} observations of our Orion sample were carried out between January 2020 and November 2021. Sixteen of these stars were observed as part of the Disk Evolution Study Through Imaging  of Nearby Young Stars (DESTINYS) program \citep{2020A&A...642A.119G, 2021ApJ...908L..25G}. The remaining seven stars are from the program 0104.C-0195(A), which targeted young intermediate-mass stars in Orion. We used SPHERE in the dual-polarization imaging (DPI, \citealt{deBoer2020, vanHolstein2020}) mode with pupil tracking enabled. All observations were conducted in the broad H-band filter (BB\_H, $\lambda_c = 1.625\mu$m.) For the majority of the sample, the science images were taken with a coronagraphic mask in place, with an inner working angle of 92.5\,mas\footnote{The inner working angle in this case refers to the radius at which the signal suppression of the coronagraph drops below 50\%.}. The exceptions are the observations of RY\,Ori, V\,1044\,Ori, and V\,2149\,Ori, which were instead taken with short individual frame exposure times (DITs) of 0.84\,s in order to prevent saturation of the primary star. We give a summary of the observation dates, setups, and observing conditions in Table~\ref{tab:obslog}. 

The data were reduced using the IRDIS Data reduction for Accurate Polarimetry\footnote{https://irdap.readthedocs.io} (IRDAP, \citealt{vanHolstein2020}) with default settings. The pipeline applied flat-field, sky, and bad-pixel corrections. It then used dedicated center calibration frames (for the coronagraphic images) to re-center all images. In the center calibration frames, a waffle pattern was introduced to the adaptive optics deformable mirror, which in turn produced equidistant calibration satellite spots around the primary star position. For the non-coronagraphic images, frames were re-centered by directly fitting a Moffat function to the central star point spread function (PSF). IRDAP then performed polarization differential imaging (PDI, \citealt{Kuhn2001}) in order to remove the (mostly) unpolarized stellar light and retain the linearly polarized, scattered light from circumstellar dust. Additionally, IRDAP also stacked all individual exposures to create a total intensity image, sensitive to (sub)stellar companions to our target stars. Finally, since the data were taken in pupil tracking mode; in other words, with the telescope pupil stabilized, it was possible to also perform angular differential imaging (ADI, \citealt{Marois2006}) in order to obtain high-contrast total intensity observations. We note however that our observations were not optimized for this mode; in other words, the range of parallactic angles sampled during the observations (and thus the amount of field rotation in the pupil stabilized mode) was limited (the range is between 3.6$^\circ$ for RY\,Ori and 54.6$^\circ$ for V\,351\,Ori.) 

The final products of our reduction are the azimuthal Stokes, $Q_\phi$, parameters \citep{vanHolstein2020}. The Q$_\phi$ image probes the centro-symmetric pattern of the polarized light, containing all the linearly polarized signal with azimuthal (positive signal) and radial (negative signal) oriented angles of linear polarization. For single scattering of dust particles illuminated by a central source, we expect all linear polarized light to be contained as a positive $Q_\phi$ signal. $Q_\phi$ is thus typically nearly identical to the corresponding polarized intensity images (created by adding the measured Stokes $Q$ and $U$ frames in quadrature), but with more favorable noise properties. For a recent definition of the $Q_\phi$ formalism in the context of circumstellar disk observations, we refer to \cite{Monnier2019}. 

\subsection{ALMA Band 6 observations}
Millimeter photometry toward the sources in our sample were obtained with ALMA in the Cycle 8 program 2021.1.01705.S (PI: C. Ginski.) The Band 6 observations were carried out between June and September 2022 with a median precipitable water vapor level of 0.5\,mm. The correlator was configured with a spectral window centered on the $J=2-1$ transition of CO, though here we concentrate only on the continuum data at a mean frequency of 241\,GHz (1.24\,mm.) Baseline lengths varied from $\sim 40-300$\,m, corresponding to a beam size $\sim 0\farcs35 - 0\farcs4$. The integration times on each source for this snapshot program ranged from 90 to 300 seconds and the median sensitivity was $70\,\mu$Jy/beam. Most sources were unresolved and fluxes were measured from the pipeline reduced data products using a $1''$ aperture. For larger sources, we used a custom aperture that encompassed all the emission above three times the rms noise level. The flux measurements are reported in Table \ref{tab:phot}. A more detailed analysis will be performed in future papers. 

\subsection{X-shooter spectroscopy}
Mid-resolution spectroscopic observations were taken with the X-shooter spectrograph \citep{2011A&A...536A.105V} mounted on the Very Large Telescope (VLT) at Paranal. The X-shooter ranges from 300-2480\,nm and the observations are taken with three spectroscopic arms: UVB (300-559.5\,nm), VIS (559.5-1024\,nm), and NIR (1024-2480\,nm.) The chosen slit widths of 1.0\arcsec, 0.4\arcsec\ and 0.4\arcsec\ resulted in a resolving power of R\,~\,5,400 in the UVB, R\,~\,18,400, in the VIS, and R\,~\,11,600 in the NIR, respectively. 
The observations were taken in nodding mode and were reduced with the X-shooter pipeline build 3.3.5 \citep{Modigliani2010}. They were flat-field- and bias-corrected, and wavelength-calibrated in this routine. Spectrophomometric standards from the ESO database were used for flux calibration.

\begin{table*}
\caption{Observing dates, conditions, and setup.}
\label{tab:obslog}      
\centering             
\begin{tabular}{l c c c c c}
\hline\hline                
Object & Date & DIT (s) & NDIT & N$_\mathrm{cycle}$ & Total integration time (min)\\
\hline

Kiso\,A-0904\,60 & 2021-02-01 & 12 & 2 & 32 & 51.2\\
Brun\,216 & 2020-01-13 & 64 & 1 & 7 & 29.9 \\
Brun\,252 & 2020-02-22 & 64 & 1 & 4 & 17.1 \\
Haro\,5-38 & 2020-12-17 & 32 & 1 & 26 & 55.5\\ 
HD\,294260 & 2020-02-18 & 64 & 1 & 7 & 29.9\\ 
HD\,294268 & 2021-01-28 & 32 & 1 & 26 & 55.5\\ 
PDS\,110 & 2020-12-10 & 32 & 1 & 26 & 55.5\\
PDS\,113 & 2021-11-03 & 32 & 1 & 26 & 55.5\\
RV\,Ori & 2021-02-06 & 32 & 1 & 26 & 55.5\\
RY\,Ori & 2020-02-24 & 0.84 & 12 & 8 & 5.4\\
TX\,Ori & 2020-11-13 & 32 & 1 & 26 & 55.5 \\
V\,351\,Ori & 2021-01-21 & 32 & 1 & 36 & 76.8\\
V\,499\,Ori & 2020-12-18 & 32 & 1 & 26 & 55.5\\
V\,543\,Ori & 2020-11-15 & 32 & 1 & 26 & 55.5\\
V\,578\,Ori & 2021-10-29 & 32 & 1 & 26 & 55.5\\
V\,599\,Ori & 2021-02-07 & 32 & 1 & 26 & 55.5\\
V\,606\,Ori & 2021-02-05 & 32 & 1 & 26 & 55.5\\
V\,1012\,Ori & 2021-02-04 & 32 & 1 & 27 & 57.6\\
V\,1044\,Ori & 2020-02-19 & 0.84 & 12 & 8 & 5.4\\
V\,1650\,Ori & 2020-02-07 & 64 & 1 & 8 & 34.1\\
V\,1787\,Ori & 2021-11-03 & 32 & 1 & 26 & 55.5\\
V\,1788\,Ori & 2021-09-27 & 32 & 1 & 27 & 57.6\\
V\,2149\,Ori & 2020-02-16 & 0.84 & 12 & 8 & 5.4\\

\hline                                  
\end{tabular}\\
\smallskip
All observations were taken in the BB\_H filter.
\end{table*}

\section{Sample and stellar properties} \label{sec:sample}
In this section, we present the sample and determine the stellar properties. 
Our target sample derives from the DESTINYS ESO large program (PI: Ginski) as well as the open time program 0104.C-0195(A) (PI: Waters.) For the initial target selection, we compiled a catalog of stars located at the distance and on-sky coordinates of Orion. We only selected targets that had some indication of youth, as recorded by Simbad (e.g., they were listed as young stellar objects, T Tauri stars, or Herbig stars.) We then filtered this catalog based on their observability with SPHERE; in other words, all targets have a Gmag $\leq 13$, so that the adaptive optics system of SPHERE could be operated. We furthermore selected targets based on their spectral energy distribution (SED) and only included systems that showed significant near or far-IR excess emission, indicative of a dusty disk present in the system. We did not take into account previously known information on multiplicity to avoid biasing our sample toward single stars. Additionally, the criterion for stars included in the open time program 0104.C-0195(A) was that abundant photometric and spectral data exists that well characterizes the stars as intermediate-mass T Tauri stars. We note that our sample is in all likelihood\footnote{The main factor for incompleteness probably arises from the Simbad criterion that we employed. Furthermore some stars may be strongly variable in the optical and their listed magnitude may not reflect accurately their potential observability with SPHERE. Some Gaia parallaxes may include effects of orbital motion of unresolved binaries which may place some systems artificially at closer distances in front of Orion in which case we may have excluded them erroneously. Finally SEDs were checked by eye without detailed fits of stellar models so that our sample is somewhat biased against systems with small near-IR excess and no mid- or far-IR excess.} far from complete and should be regarded in a sense as a pilot study of Orion, demonstrating the detectability of extended circumstellar disks at distances up to 420\,pc. \\
As is clear from Table \ref{tab:stars}, our stars span from early-A (V1788 Ori) to late-K (V578 Ori) spectral types. Based on $Gaia$ DR3, their distances cover most of the Orion depth, spanning the range from 320 pc (V351 Ori) to 415 pc (TX Ori.) Their spatial distribution within Orion is described in Sect.\,\ref{sec:results}. The disks in the sample that lacked previous ALMA data were also selected for additional ALMA observations with the aim of obtaining a flux measurement.

In the following, we describe the characterization of the stellar parameters. We used X-shooter spectra to determine the temperature and luminosity, as is described in section \ref{sec-tstar-lstar}. For stars without X-shooter spectra or the two cases where nonphysical values were obtained with the method we used, we instead adopted stellar parameters from the literature. This can for example occur due to stellar variability and photometry taken at different dates or to unusually high local reddening where the assumed $R_{v}$=3.1 is too low, leading to a low luminosity estimate that puts the star under the main sequence.

\renewcommand{\arraystretch}{1.5}
\begin{table*} 
\caption{Stellar properties of the sample. }
\label{tab:stars}\tiny  
\centering             
\begin{tabular}{lllllll}
Name                        &        d [pc]                     & SpT &          $T_{\rm eff}$        &       $L_{\odot}$             &        $M_{\odot}$            &   $t$ [Myr] \\\hline
Kiso A-0904 06 A$^1$        &        $339.78^{+1.74}_{-1.84}$   & K6  &          $4205^{+235}_{-105}$ &       $0.45^{+0.08}_{-0.02}$ &        $0.82^{+0.18}_{-0.13}$ &   $6.66^{+5.55}_{-3.14}$ \\
Kiso A-0904 06 B$^1$        &        $339.78^{+1.74}_{-1.84}$   & K6  &          $4205^{+235}_{-105}$ &       $0.45^{+0.08}_{-0.02}$ &        $0.82^{+0.18}_{-0.13}$ &   $6.66^{+5.55}_{-3.14}$ \\
Brun 216$^2$                &        $383.52^{+3.36}_{-2.52}$   & F8  &          $6170^{+80}_{-160}$  &       $9.55^{+2.20}_{-1.04}$  &        $1.80^{+0.25}_{-0.09}$ &   $4.46^{+0.25}_{-1.47}$ \\
Brun 252$^3$                &        $383.64^{+6.66}_{-6.43}$   & G1  &          $5890^{+120}_{-120}$ &       $7.16^{+2.51}_{-0.38}$  &        $1.68^{+0.25}_{-0.09}$ &   $6.30^{+1.24}_{-2.18}$ \\
Haro 5-38                   &        $400.35^{+2.92}_{-2.58}$   & K7  &          $4050^{+250}_{-60}$  &       $1.19^{+0.01}_{-0.06}$  &        $0.58^{+0.31}_{-0.05}$ &   $1.01^{+0.91}_{-0.11}$ \\
HD 294260                   &        $387.44^{+3.03}_{-2.42}$   & G1  &          $5880^{+50}_{-110}$  &       $5.59^{+0.40}_{-0.37}$  &        $1.53^{+0.13}_{-0.03}$ &   $7.93^{+0.80}_{-1.70}$ \\
HD 294268                   &        $361.98^{+1.98}_{-1.48}$   & G0  &          $5970^{+80}_{-40}$   &       $5.67^{+1.29}_{-0.27}$  &        $1.49^{+0.15}_{-0.02}$ &   $8.84^{+0.51}_{-2.21}$ \\
PDS 110                     &        $343.66^{+2.82}_{-2.75}$   & F6  &          $6340^{+210}_{-60}$  &       $6.70^{+0.87}_{-0.18}$  &        $1.46^{+0.10}_{-0.04}$ &   $9.82^{+0.73}_{-1.47}$ \\
PDS 113$^4$                 &        $352.25^{+2.33}_{-2.27}$   & F3+F4 &        $6750^{+70}_{-80}$  &        $3.75^{+0.82}_{-0.38}$  &        $1.32^{+0.05}_{-0.06}$ &   $16.67^{+0.54}_{-1.41}$ \\
RV Ori                      &        $397.54^{+2.44}_{-2.41}$   & K5  &          $4450^{+150}_{-150}$ &       $0.57^{+0.03}_{-0.11}$  &        $0.95^{+0.06}_{-0.04}$ &   $8.29^{+8.40}_{-2.82}$ \\
RY Ori                      &        $346.82^{+2.29}_{-2.44}$   & F5  &          $6510^{+160}_{-160}$ &       $7.17^{+0.61}_{-0.82}$  &        $1.45^{+0.11}_{-0.03}$ &   $10.12^{+0.90}_{-1.65}$ \\
TX Ori                      &        $415.53^{+37.91}_{-33.04}$ & K4  &          $4620^{+210}_{-220}$ &       $3.57^{+0.02}_{-0.45}$  &        $1.30^{+0.35}_{-0.36}$ &   $0.98^{+1.06}_{-0.40}$ \\
V351 Ori                    &        $323.82^{+3.14}_{-3.09}$   & A9  &          $7440^{+150}_{-220}$ &       $25.49^{+1.50}_{-1.54}$ &        $2.08^{+0.05}_{-0.02}$ &   $4.20^{+0.49}_{-0.25}$ \\
V499 Ori                    &        $331.61^{+1.59}_{-1.48}$   & K6  &          $4200^{+240}_{-100}$ &       $0.36^{+0.06}_{-0.05}$  &        $0.81^{+0.19}_{-0.11}$ &   $9.84^{+3.90}_{-4.65}$ \\
V543 Ori$^1$                &        $338.39^{+1.84}_{-2.09}$   & K4  &          $4590^{+240}_{-150}$ &       $0.36^{+0.36}_{-0.14}$ &        $0.82^{+0.23}_{-0.10}$ &   $22.90^{+22.18}_{-16.38}$ \\
V578 Ori                    &        $375.21^{+2.02}_{-1.97}$   & K7  &          $4050^{+250}_{-60}$  &       $1.08^{+0.32}_{-0.08}$  &        $0.59^{+0.31}_{-0.07}$ &   $1.16^{+1.11}_{-0.43}$ \\
V599 Ori$^2$                &        $401.07^{+4.06}_{-3.27}$   & A5  &          $8000^{+250}_{-250}$ &       $27.54^{+1.15}_{-1.15}$ &        $2.03^{+0.10}_{-0.10}$ &   $4.29^{+0.42}_{-0.54}$ \\
V606 Ori                    &        $400.14^{+3.66}_{-2.86}$   & K5  &          $4450^{+150}_{-150}$ &       $0.67^{+0.03}_{-0.13}$  &        $1.03^{+0.06}_{-0.13}$ &   $6.47^{+6.98}_{-2.33}$ \\
V1012 Ori$^2$               &        $376.23^{+6.60}_{-5.79}$   & A3  &          $8500^{+250}_{-250}$ &       $5.89^{+0.72}_{-0.64}$  &        $1.30^{+0.07}_{-0.07}$ &   $8.47^{+1.06}_{-0.89}$ \\
V1044 Ori                   &        $384.31^{+2.70}_{-2.77}$   & G6  &          $5590^{+70}_{-40}$   &       $4.64^{+0.23}_{-0.09}$  &        $1.64^{+0.06}_{-0.02}$ &   $6.00^{+0.36}_{-0.59}$ \\
V1650 Ori$^3$               &        $342.41^{+5.63}_{-5.46}$   & F8  &          $6160^{+100}_{-100}$ &       $9.53^{+0.25}_{-0.78}$  &        $1.71^{+0.05}_{-0.07}$ &   $6.20^{+0.96}_{-0.41}$ \\
V1787 Ori                   &        $393.08^{+2.86}_{-2.73}$   & A5  &          $8150^{+185}_{-160}$ &       $14.13^{+4.07}_{-2.64}$ &        $1.66^{+0.09}_{-0.08}$ &   $7.43^{+0.59}_{-1.05}$ \\
V1788 Ori                   &        $347.65^{+4.98}_{-5.95}$   & A0  &          $9700^{+1000}_{-400}$&       $19.06^{+9.19}_{-2.89}$ &        $2.20^{+0.08}_{-0.20}$ &   $5.08^{+3.17}_{-0.11}$ \\
V2149 Ori A$^3$             &        $388^{+5}_{-5}$            & F8  &          $6180^{+110}_{-110}$ &       $17.9^{+2.80}_{-0.45}$ &        $2.1^{+0.16}_{-0.07}$ &   $3.57^{+0.14}_{-0.99}$ \\
V2149 Ori B$^3$             &        $388^{+5}_{-5}$            & F8  &          $6180^{+110}_{-110}$ &       $17.9^{+2.80}_{-0.45}$ &        $2.1^{+0.16}_{-0.07}$ &   $3.57^{+0.14}_{-0.99}$ \\
\hline                                  
\end{tabular}\\
\smallskip
\normalsize
\raggedright
$^{1} T_{\rm eff}$ is taken from \citet{2019AJ....157...85B}. $^2$ All values from \citet{2018A&A...620A.128V}. $^3$ Values from \citet{2021A&A...652A.133V}.  $^4$ Spectroscopic binary (SB2) \citet{2008AstBu..63..272K}; we detect both components in the X-shooter spectra. Stellar parameters are given for one component, assuming the stars have roughly equal luminosity.
The method used to obtain the stellar parameters is described in Sect.\,\ref{sec:sample}, unless differently indicated in the footnotes. The columns are: source name, distance, effective temperature, stellar luminosity, mass, and age.\\
\end{table*}

\subsection{Stellar temperature and luminosity}
\label{sec-tstar-lstar}
The temperatures of the stars were determined by spectrally classifying the stars with the use of the PyHammer code \citep{Roulston2020} and a spectrum comparison with Gray's spectral classification atlas \citep{gray2000digital}. The first guess of spectral classes was done with the PyHammer code, a tool to automatically classify stars. The first guess from PyHammer was used as a preselection before comparing the spectral features of the spectra with the spectral classification atlas from Gray. Comparing the ranges of spectral classes and luminosity classes, and comparing the stars among each other, results in a spectral classification for each star. The temperatures corresponding to the derived spectral class were taken from Tables 5 and 6 of \cite{2013ApJS..208....9P}. For the stars where we lacked X-shooter spectra, we did the following: for Brun 252, V1650 Ori, and V2149 Ori we adopted the stellar parameters from \citet{2021A&A...652A.133V}, for Kiso A-0904 60 and V543 Ori we adopted the spectral class from \citet{2005AJ....129..907B}, and for Brun 216 we adopted the stellar parameters from \citet{2018A&A...620A.128V}.  

The luminosity was determined by first estimating the extinction, $A_{V}$, by the method described in \citet{2021A&A...652A.133V} (section 2.3), using the effective temperature table for standard stars in \citet{2013ApJS..208....9P} and B- and V-band photometry from the NOMAD \citep{2004AAS...205.4815Z} and UCAC4 photometric catalogs \citep{2013AJ....145...44Z}. To obtain the uncertainty in luminosity, we repeated this process but varied the temperature within its obtained uncertainty range. We give the results in Table~\ref{tab:stars}. In the cases of V599 Ori and V1012 Ori, our method gives spurious locations in the Hertzsprung-Russell (HR) diagram. For V1012, the luminosity was difficult to estimate due the highly inclined disk (section \ref{sec:V1012}) effectively obscuring the stellar light. For V599 Ori, the method suggested an age comparable to a main-sequence star that would have already lost its disk. With the high-quality X-shooter spectra, this suggests that the luminosity estimate in the method is too low. One reason for this could be a high local extinction coming from material close to the star. For these two systems we therefore adopted the stellar parameters from \citet{2018A&A...620A.128V}.

\begin{figure*}
        \includegraphics[width=\textwidth]{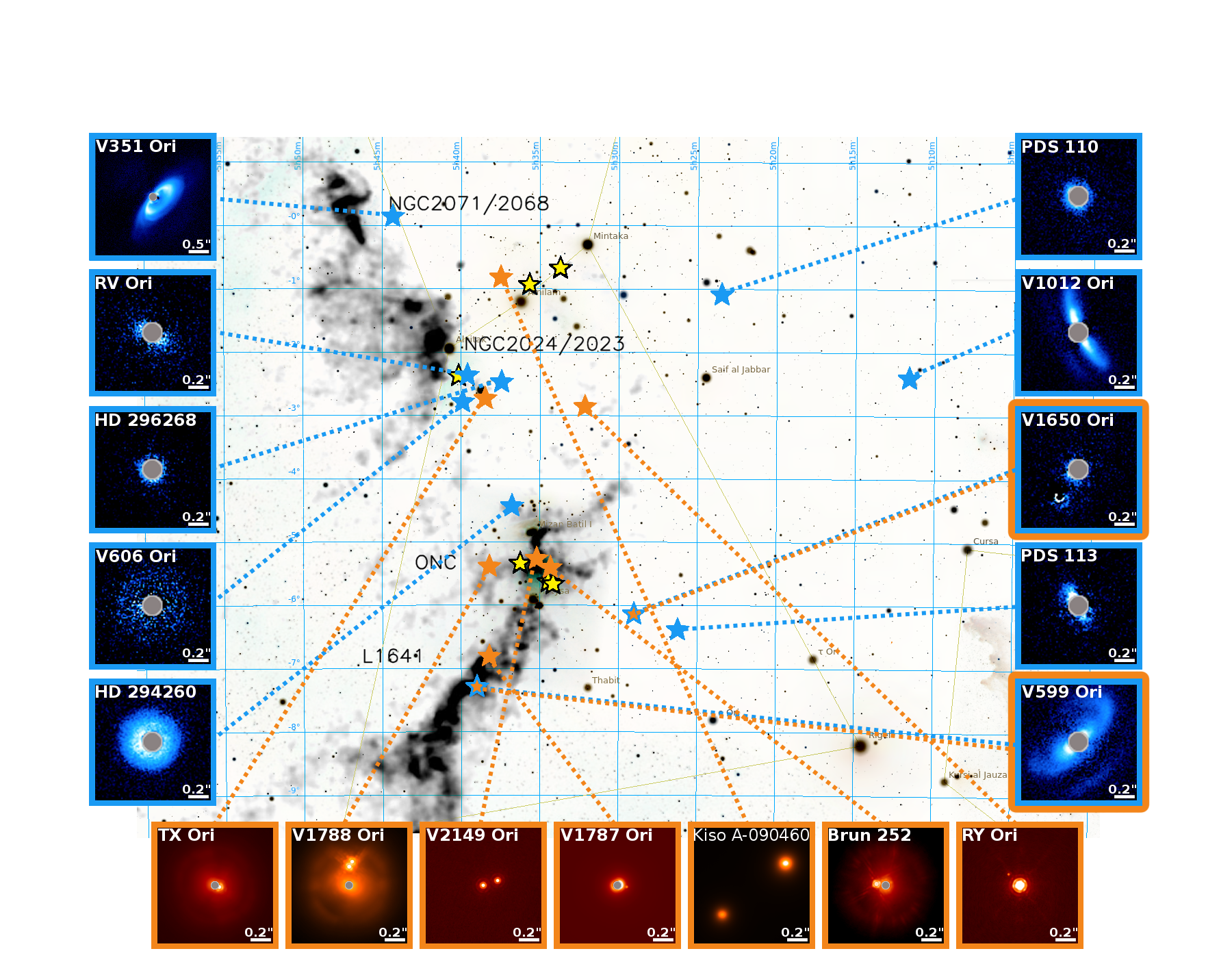}
        \caption{Imagery of the sample. The figure shows a stellar map with the location of the sources in this paper. The thumbnails show disk detections (blue), detected companions (orange), and non-disk detections (yellow.) The extent of the coronagraph is displayed as a gray circle in each thumbnail when used. The stellar map was generated with Stellarium (open source software https://stellarium.org/) and the IR extinction map (dark gray clouds) was adapted from \citet{2012AJ....144..192M} based on \citet{2011ApJ...739...84G}.}
        \label{fig:masteroforion}
    \end{figure*}

\subsection{Stellar masses and ages}

The stellar mass and the stellar age were derived using \citet{2000A&A...358..593S} PMS evolution models. Uncertainties on the age and mass were obtained by propagating the uncertainty in temperature and luminosity. This means that for low-mass stars, where the evolutionary tracks are close to one another in the HR diagram, small changes in temperature and luminosity can have a relatively large effect on both mass and age. For high-mass stars, isochrones have more spacing and lead to smaller error bars on these quantities. This is a well-known issue when determining stellar parameters for low-mass stars,  independently of the specific evolutionary models that are being used \citep{2016MNRAS.461..794P}. We list the final results along with the stellar temperature and luminosity in Table ~\ref{tab:stars}.

\section{Results from SPHERE} \label{sec:results}
 

Figure \ref{fig:masteroforion} shows the spatial distribution within Orion of our sample, and we show the images revealing disk detections (from the $Q_\phi$ images in the blue panels) as well as those revealing the presence of any companion candidates (from the intensity images in the orange panels.) Interestingly, the two categories appear to be mutually exclusive in our sample. In other words, no companions are detected then the disk is resolved, and vice versa. In this section, we first describe the detection of stellar companions from the SPHERE image, and then move to the characterization of the disk signal.

\subsection{Multiplicity}
Using the total intensity H-band observations obtained with SPHERE, we detected close (sub)stellar companions\footnote{We refer to the detected sources as companions, but note that strictly these are companion \textit{candidates} since in principle it is possible that these are background objects in chance-alignment. However, given their small projected separation ($<$1") to the primary star in all cases and the fact that these observed fields are not crowded, the probability for this scenario is low.} in nine of the 23 systems in our sample. Of these nine systems, three were previously known in the literature, namely the triple system V1788 Ori \citep{2007IAUS..240..250T} and the two double systems V2149 Ori \citep{2006A&A...458..461K, 2009ApJ...697.1103T} and Kiso A-090460 \citep{2020AJ....160..268T}. The remaining six systems with companions are new detections. In the following, we first describe the astrometric and photometric extraction and then briefly report on the individual systems.

\subsubsection{Astrometric and photometric extraction}
Since the majority of the detected companions do not show a high contrast relative to the central star, we used the non-coronagraphic flux calibration frames when possible to extract the primary and companion positions from the same frame. Flux calibration images were taken for the entire target sample and consist of short exposures, in some cases with a neutral density filter inserted to avoid saturation of the primary star.

The astrometric extraction was then performed by simultaneously fitting two Gaussian PSFs to the total intensity image. The fitted Gaussians were then subtracted from the data and the residuals were examined to confirm that a good fit had been obtained. The location of the peak of the Gaussians was taken as the center of the stellar PSF. For two targets (TX Ori and V1787 Ori), an accurate determination of the location of the companion was hindered by the strong glare of the primary component. In these two cases, we fitted the primary and secondary positions separately following the approach used in \citet{2013MNRAS.434..671G,2018A&A...616A..79G}. We therefore first determined the primary position by fitting a single Gaussian. We then rotated the frame by 180$^\circ$ around this position and subtracted the rotated frame from the original. In this way, the radially symmetric component of the primary star PSF was removed without affecting the signal of faint nearby companions. We then fitted a second Gaussian to the companion position in the subtracted frame to obtain its position.
All the results can be found in table \ref{tab:multiple_data}.

To extract the photometric data from the total intensity frames, we used a circular aperture with a radius of 4 pixels. The apertures were centered on the locations obtained from the astrometric extraction. To accurately measure the fainter companion star, we modeled the background flux at the companion positions, which was in all cases dominated by the wings of the PSF of the bright stellar primary. 

Assuming that the diffuse emission of the PSF wings is to the first order point-symmetric, we measured the flux 180$^{\circ}$ from the position angle (PA) of the secondary and with the same separation. To capture the local background close to the companion position, we then took eight additional measurements at the same separation, four in the clockwise direction and four in the counterclockwise direction, in increments of 20$^{\circ}$ from the companion. From these nine measurements, the local average intensity was calculated and also removed from the measurement of the companion. The uncertainty in the measurement was estimated using the standard deviation of these nine background measurements. From these measurements, the contrast between the primary and the companion was calculated.    

\subsubsection{Description of the systems}
The masses and spectral classes of the stellar companions were derived using the \citet{2000A&A...358..593S} pre-main-sequence stellar evolution models, assuming that the age and distance of the companion is the same as the age and distance of the primary. We used 2MASS H-band magnitudes \citep{2003yCat.2246....0C} for the primary component and our measured H-band contrast to calculate the absolute magnitude in the H band for the companions via the distance modulus. Since the 2MASS H-band measurement is a combination of the components in the system, we adopted the magnitude adjustment of the primary described in \citet{2020A&A...635A..73B} to account for this. The H-band reddening was obtained by converting the A$_{V}$ into A$_{H}$ using the conversion described by \citet{1989ApJ...345..245C}. Here, we briefly discuss the individual cases. 

\begin{table*}
\caption{Values from the photometric and astrometric extraction together with the estimated spectral class and mass of companions.}
\label{tab:multiple_data}      
\centering             
\begin{tabular}{l c c r r r }     
\hline\hline                
Object & Derived SpC & Mass M$_{\odot}$ & Sep (arcsec) & PA ($^\circ$) & $\Delta$mag \\
 & & & $\pm{0.01}$ & $\pm{0.48}$ & \\
\hline
\hline
Kiso\,A-0904\,60    & K6    & 0.8       & 1.96  & 128.95    & 0.75\\
Brun\,252           & M0    & 0.6       & 0.22  & 82.42     & 2.69\\
RY\,Ori             &       & 0.076     & 0.39  & 45.57     & 6.26\\
TX\,Ori             & M5    & 0.2       & 0.11  & 248.84    & 2.14\\
V\,599\,Ori         & M6    & 0.1       & 2.05  & 40.40     & 5.59\\
V\,1650\,Ori        & K7    & 0.8       & 0.32  & 144.09    & 2.83\\
V\,1787\,Ori        & M3    & 0.2       & 0.22  & 189.93    & 5.22\\
V\,1788\,Ori\,(b)   & K6    & 0.9       & 0.43  & 1.65      & 2.47\\
V\,1788\,Ori\,(c)   & K7    & 0.7       & 0.56  & 0.49      & 2.84\\
V\,2149\,Ori        & F8    & 2.1       & 0.37  & 288.34    & 0.58\\
\hline                                  
\end{tabular}
\end{table*} 

\textit{Kiso A-090460}. The wide binary companion at PA = 128.95$^{\circ}$ has a separation of 1.96\arcsec, which agrees with previous measurements for the separation \citep{2020AJ....160..268T}. The H-band magnitude contrast is 0.75\,mag and corresponds to an absolute magnitude of 3.75\,mag. This classifies as an K6 star of mass 0.8 M$_{\odot}$. Considering the low H-band contrast and the mass of the companion, this is most likely an equal-mass binary.

\textit{Brun 252}. In this observation, we detect a companion at PA = 82.42$^{\circ}$ and with a separation of 0.22\arcsec (84\,au.) The magnitude contrast in the H band to the primary is 2.69\,mag, which corresponds to an absolute magnitude in H band of 4.01\,mag and a M0 star of mass 0.6 M$_{\odot}$. 


\textit{TX Ori}. A close companion is visible in the west at PA = 248.84$^{\circ}$ with a separation of 0.11\arcsec. The companion star has a magnitude contrast in the H band of 2.14\,mag, which corresponds to an absolute H-band magnitude of 3.31\,mag and a M5 star of 0.2 M$_{\odot}$.

\textit{V599 Ori}. A wide companion is detected in the frame of V599 Ori. It is located at PA = 40.14$^{\circ}$ with a separation of 2.04\arcsec. The contrast is measured to 5.59\, mag, corresponding to an absolute H-band magnitude of 5.71\,mag. This corresponds to a M6 star with a mass around 0.1 $M_{\odot}$. In Sect.\,\ref{sec:discussion} we evaluate the probability of the companion being bound.

\textit{V1650 Ori}. We detect a new stellar companion in the system of V1650 Ori. It is located at PA = 144.09$^{\circ}$ with a separation of 0.32\arcsec to the primary. The contrast is 2.83\,mag corresponding to an absolute H-band magnitude of 3.51\,mag. Using Siess stellar isochrones, this corresponds to a K7 star with a mass of around 0.8 $M_{\odot}$ at the same age. 

\textit{V1787 Ori}. In the case of V\,1787\,Ori a new companion is located at a separation of 0.22\arcsec\ (87\,au) and has a magnitude contrast of 5.2\,mag. This corresponds to an absolute magnitude of 5.61\,mag in the H band, considering a distance of 393.1\,pc. This corresponds to an M3 red dwarf of mass 0.2 M$_{\odot}$. We also detect the possible companion V1787 Ori B \citep{2021MNRAS.501.1243A} just at the very edge of our frame. The separation is estimated as 6.6\arcsec and at PA$\sim 18^{\circ}$ but the closeness to the edge of the frame makes it difficult to use our method for astrometric extraction.

\textit{V1788 Ori}. We detect both previously known companions in V1788 Ori. The secondary at PA = 358.35$^{\circ}$ has a separation of 0.43\arcsec and the tertiary at PA = 359.51$^{\circ}$ has a separation of 0.56\arcsec. The H-band magnitude contrast we measured as 2.47\,mag and 2.84\,mag, respectively. This corresponds to spectral class K6 and a mass of 0.9 M$_{\odot}$ for the secondary component and spectral class K7 and a mass of 0.7 M$_{\odot}$ for the tertiary one. 

\textit{V2149 Ori}. The separation of the known binary V2149 Ori was measured as 0.33$\pm{0.01}$\,arsec in December 2001 by \citet{2006A&A...458..461K}. We find a measurable motion of the secondary, with the new measurement of the separation being 0.365$\pm{0.016}$\arcsec. This corresponds to a relative velocity of  $\sim3.6 \rm kms^{-1}$ at a distance of 388\,pc. This velocity is not large enough to exclude a bound orbit. The secondary is located at PA = 288.39$^{\circ}$ and has an H-band magnitude contrast of 0.58\,mag, corresponding to an absolute H-band magnitude of -0.29\,mag. Using the Siess evolutionary tracks, we find that the companion is an F8 star of approximately 2.1-2.2\,M$_{\odot}$. Considering the low H-band contrast and the mass of the companion, this is most likely an equal-mass binary.

\subsubsection{Possible substellar companion of RY Ori}

Around one of the systems in this study, RY\,Ori, we detected a close companion that is particularly faint. It is located at a projected separation of 0.39\,mas, corresponding to 137\,au at a distance of 350.5\,pc. Using aperture photometry as was previously described, we find a magnitude contrast of 6.14\,mag between the primary star and the companion in the H band. This corresponds to an absolute magnitude of 7.34\,mag. Since this is faint for a stellar companion, we used in this case DUSTY evolutionary tracks (\citealt{Chabrier2000}) to derive the mass of the object. We adopt an age of 10.1\,Myr, and find a mass of 80\,M$_{\rm Jup}$ for this object; that is, at the low end of the stellar mass regime. If we instead assume the lower end of the age uncertainty, we derive for the system (8.4\,Myr), then we rather find a mass of $\sim$72\,M$_{\rm Jup}$; that is, this would place the object at the boundary of the stellar-substellar regime. Spectroscopic observations of the object are required for a detailed characterization.  \\

\subsection{Circumstellar disks}
\begin{figure*}
        \includegraphics[width=\textwidth]{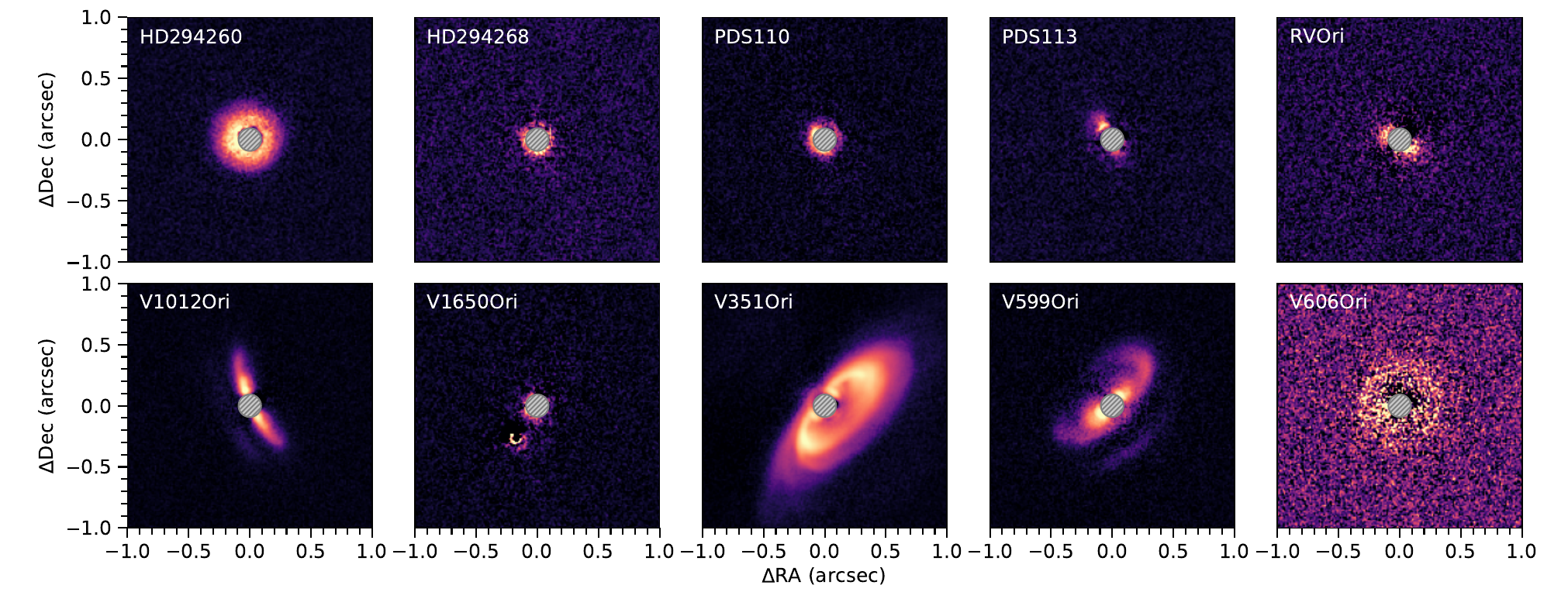}
        \caption{All disks detected in our SPHERE H-band data. We show the Q$_\phi$ (azimuthally) polarized light images in all cases. All systems are shown on a logarithmic color map, individually adjusted to highlight the disk morphology. The hashed gray circle in the center of the images indicates the position and size of the coronagraphic mask (centered on the system's primary star.)}
        \label{fig:diskgal}
    \end{figure*}

Using the $Q_\phi$ images, we determined the presence of circumstellar disks in ten sources. These are all shown in Fig.\,\ref{fig:diskgal}. For all disks, we measure the inclination, PA, and extent of detectable emission as is described in Appendix \ref{sec:disk_measurement}. All the measured disk parameters are listed in Table \ref{tab:geometric_data}. 

To evaluate the actual disk brightness, we used the polarized-to-stellar light contrast, $\delta_{\rm pol}$ \citep{Garufi2017, 2023ASPC..534..605B}. This contrast is a measurement of the fraction of photons released by the star that are effectively scattered toward the observed, and depends on several factors but primarily on the presence of shadows from the inner disk regions, the disk flaring angle, and the dust properties. We classify disks with $\delta_{\rm pol}\leq 3$ as faint, similar to \citet{2022A&A...658A.137G}, and disks with $\delta_{\rm pol}>3$ as bright. As a result, our sample contains three bright disks (V351 Ori, V599 Ori, and V1012 Ori), seven faint disks, and thirteen non-detections that are all discussed in the following. To highlight the asymmetries in the bright disks we created contrast images (Fig. \ref{fig:diffdisk}) by flipping the original frame along the major axis of the disk and dividing the original frame with the flipped frame. In this way we highlight the relative intensities in the asymmetries. 

\subsubsection{Bright disk of V351 Ori}

The disk surrounding V\,351\,Ori is the most extended disk in our sample, with an outer radius of 1.12\arcsec (361\,au) along the major axis. This system was first observed by \cite{wagner2020first} who detected the disk in total intensity L-band observations using angular differential imaging. In their observations, they find a bright and possibly asymmetric ring at 0.4\arcsec (130\,au\footnote{This value is based on the distance of 323.8\,pc that we adopt for this source. In \cite{wagner2020first} this corresponds to 140\,au.}.) Our new polarized light observations in the H band show a wealth of additional substructures. We also detect the bright ring (a, figure \ref{fig:diffdisk}) seen by \cite{wagner2020first} and can trace the polarized scattered light signal all the way down to the coronagraphic mask (92.5\,mas, 30\,au.) 

Outside of this ring, we find that the disk is asymmetric between the southeast and the northwest. We highlight this in figure.\,\ref{fig:diffdisk}, where we show that the received scattered light signal differs by up to a factor of five along the disk's major axis. The structures outside of the bright inner ring may either be ring-like in nature or may trace a large spiral arm seen under an inclination. The structure visible at 0.332\arcsec ($\sim107$\,au) appears at first glance to trace a ring, best visible in the southwest (in the direction of the minor axis) and the northwest b, figure \ref{fig:diffdisk}, along the major axis). If this is true, then this ring shows a clear offset from the stellar position along the major axis, indicating that it is either eccentric or that the disk is warped, and thus the PA of the major axis of the outer ring (and possibly its inclination) is misaligned relative to the inner ring. 

On the other hand, two faint arc-like structures are visible from 0.700\arcsec to 1.120\arcsec in the southeast, outside the outer ring structure (c, figure \ref{fig:diffdisk}). These might be interpreted as the trailing ends of an extended spiral structure. If this is the case, then they may well connect to the previously mentioned ring-like structure interior to their position, extending the spiral closer to the star. Both the warp and the spiral interpretation may explain the asymmetry along the major axis seen in figure~\ref{fig:diffdisk}. As the detailed modeling of individual systems is beyond the scope of this study, we refer to Ginski et al. (in prep.), in which we discuss the V351 Ori system in detail. Further ALMA CO emission line observations will also be presented by Stapper et al. (in prep..) Since the scattered light disk structure is complex and the disk is resolved in our ancillary ALMA dust continuum observations, we retrieved the inclination and PA of the system from the ALMA data by fitting an axisymmetric model to the visibileties using \texttt{FrankFitter}\citep{2020MNRAS.495.3209J}. We obtain the inclination 63.2$^\circ{}$ and PA 325.8$^\circ{}$.

\begin{table*}
\caption{Results of geometric fit to detected scattered light disks. }
\label{tab:geometric_data}      
\centering             
\begin{tabular}{l c c r r c c c c}     
\hline\hline                
Object & i ($^\circ$) & PA ($^\circ$) & $r$ (mas) & $r$ (au) & offset & offset angle ($^\circ$) & $\delta_{\rm pol}$\\
& & & & &  RA/Dec (mas)& & \\
\hline

HD\,294260      & $21.68 \pm{3.09}$     & $118.64 \pm{11.22}$   &$389.01 \pm{5.01}$     &$151 \pm{2}$   &-13.94/18.23   & $37.41$   & $0.86\pm{0.13}$\\
HD\,294268      & $35.93 \pm{8.02}$     & $204.22 \pm{18.87}$   &$174.23 \pm{14.12}$    &$63 \pm{5}$    &-8.64/-25.23   & $161.09$  &$0.19\pm{0.09}$\\ 
PDS\,110        & $21.36 \pm{3.31}$     & $203.50 \pm{13.09}$   &$ 187.91 \pm{2.63}$    &$65\pm{1}$     &-13.75/-2.47   & $100.21$  &$0.26\pm{0.09}$\\ 
PDS\,113        & $60.31 \pm{7.31}$     & $214.60 \pm{9.60}$    &$ 252.39 \pm{18.37}$   &$ 89\pm{6}$    &-38.21/24.44   & $57.40$   &$0.64\pm{0.14}$\\ 
RV\,Ori         & $56.08 \pm{10.29}$    & $244.23 \pm{27.28}$   &$ 183.84 \pm{29.35}$   &$ 73 \pm{12}$  &-3.32/-19.26   & $170.23$  & $1.97\pm{0.87}$\\ 
V\,351\,Ori$^{1}$     & $63.2$          & $320.16$         &$1120 \pm{60}$         &$361\pm{20}$   & -  &     -      & $5.65\pm{0.80}$\\
V\,599\,Ori     & $56.86 \pm{4.02}$     & $137.10 \pm{4.29}$    &$ 551.16 \pm{21.07}$   &$ 221\pm{9}$   &-53.67/85.18   & $32.22$   & $3.51\pm{0.53}$\\ 
V\,606\,Ori$^{2}$     & $25  \pm{5}$         & $240 \pm{10}$           & $380\pm{20}$        &$152\pm{8}$      & - &     -      &$1.03\pm{1.00}$\\
V\,1012\,Ori    & $70.05 \pm{5.55}$     & $204.61 \pm{4.31}$    &$ 473.63\pm{37.71}$    &$ 178\pm{15}$  &63.24/44.63   & $305.21$  &$3.92\pm{0.82}$\\ 
V\,1650\,Ori    & $30.47 \pm{8.03}$     & $224.75 \pm{27.73}$   &$ 161.81\pm{11.66}$    &$55\pm{4}$     &-16.03/-27.30   & $149.59$  &$0.1\pm{0.1}$\\ 
\hline                                  
\end{tabular}
$^{1}$ For V351\,Ori we report the inclination and PA retrieved from the ALMA millimeter observations. Radius is estimated from the Q$_{\phi}$ image.$^{2}$ We note that for the weak disk V606 Ori the radius and PA were estimated by eye (see section \ref{sec:results}).
\end{table*} 

\subsubsection{Bright disk of V599 Ori}
The disk is observed at an inclination of 57$^{\circ}$ and a PA of 137$^{\circ}$. The signal extends 551\,mas ($\sim 220$\,au.) The forward scattering edge of the bottom side of the disk is faint but clearly detected (d, figure \ref{fig:diffdisk}). Using the bottom side signal, an estimate of the disk thickness can be made. This suggests a scattering surface height of $\sim140$\,au at the outer radius of the disk. The inner disk is brighter in the southeast, while the outer disk is brighter in the northwest (I and II respectively, figure \ref{fig:diffdisk}). At that location, a possible spiral arm extends from the northwest toward the north with a brightening in intensity at the base where it attaches to the rim of the scattered light disk (e, figure \ref{fig:diffdisk}). 

In the direction of the spiral arm (PA$\sim47^{\circ}$) at a distance of 2", a fainter star can be seen (f, figure \ref{fig:diffdisk}). It is not clear if this is a field star or whether it is dynamically connected to the V599 Ori system. The projected distance would be about 800\,au. If the star and the disk are at the same distance from us, the star could be interacting dynamically with the disk. The star was detected by \textit{Gaia}, but unfortunately does not have a measured parallax in the \textit{Gaia} DR3 database. In Sect.\,\ref{sec:discussion} we evaluate the possibility of the star being a companion, and thus the possibility of it being a perturber. Another interpretation of the spiral arm feature could be that a misaligned inner disk or a warp in the disk is creating an asymmetry in a ring. In that case, the shadow of the inner disk is falling on the upper, outer disk in the southeast, creating the appearance of a spiral arm.   

\subsubsection{Bright disk of V1012 Ori}\label{sec:V1012}
\begin{figure*}[ht]
        \includegraphics[width=\textwidth]{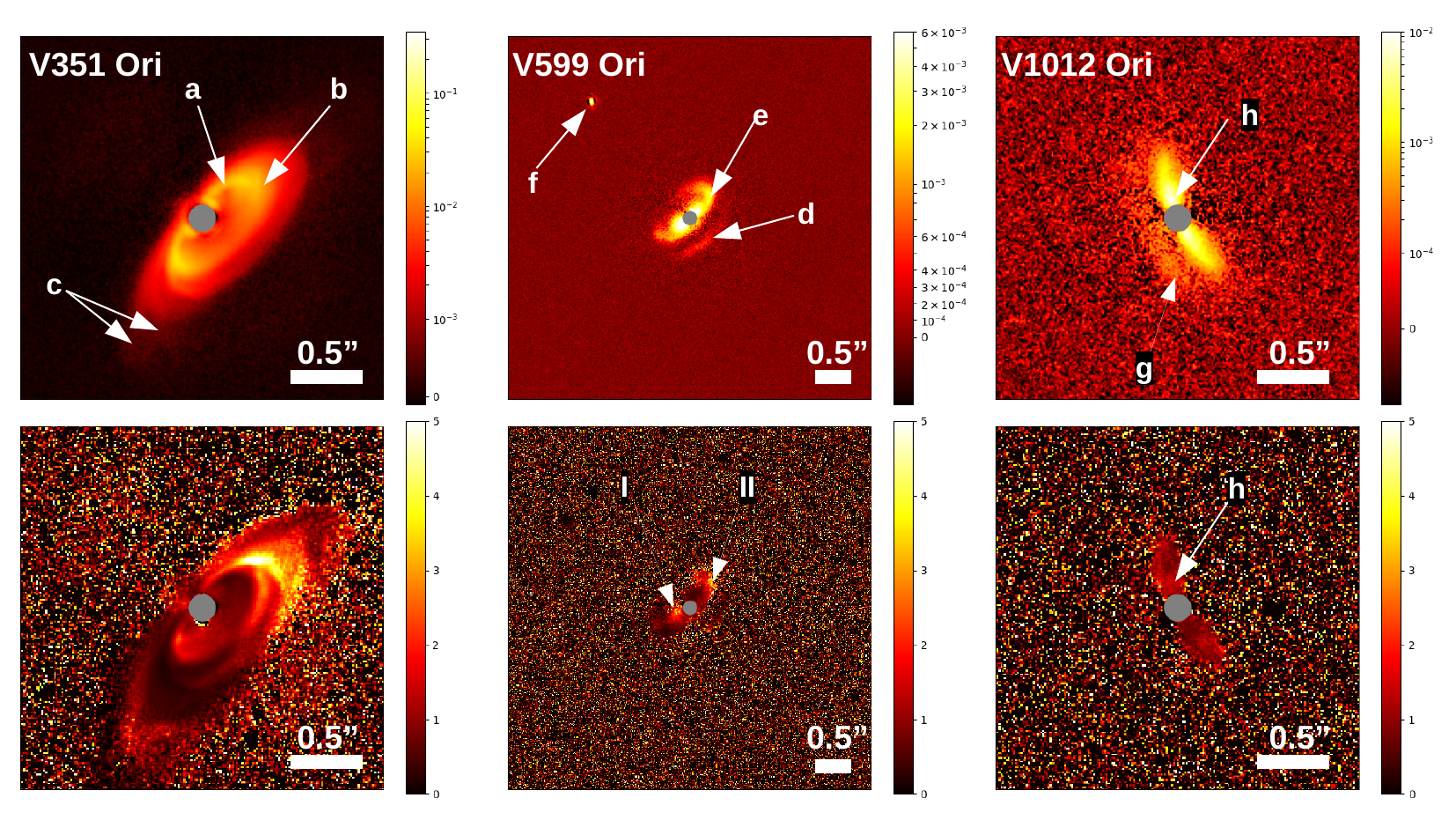}
        \caption{Images showing the asymmetry of the three bright ($\delta_{\rm pol}>3$) large disks in the sample. \textbf{Top row:} From left to right, the H-band observations of V351 Ori, V599 Ori, and V1012 Ori. \textbf{Bottom row:} Images showing the asymmetry in the disk. These were created by dividing the original frame by a frame with the inverted disk along its minor axis. From left to right, V351 Ori, V599 Ori, and V1012 Ori. }
        \label{fig:diffdisk}
    \end{figure*}

This disk has the highest inclination in the sample, 70$^{\circ}$, which means we see it close to edge on, with a PA of $\sim205^{\circ}$. The upper scattering surface is clearly visible. There is an asymmetry in the disk, with an increase in intensity in the northeast part of the disk. The lower scattering surface is faint but detectable (g, figure \ref{fig:diffdisk}). It is brighter in the southwest than in the northeast (g and h, respectively, figure \ref{fig:diffdisk}). The asymmetry in the surface brightness of the upper and lower scattering surfaces could indicate a misalignment or warp in the inner disk, which can give rise to shadows and intensity asymmetries in the disk \citep{2020A&A...636A.116K,2020ApJ...888....7L,2022A&A...658A.183B}. The scattered light radius is 474\,mas, which corresponds  $\sim178$\,au. There are no resolved structures in the disk. An estimate of the thickness of the disk can be done in the same way as in V559 Ori above and gives a scattering surface height of $\sim115$\,au at a radius of $\sim178$\,au.




\subsubsection{Faint disks}
The observed extent of the faint disks ranges from the smallest disk, V1650 Ori, with a radius $\sim162$\,mas (55\,au, only a marginal detection), to HD 294260, with a radius of $\sim389$\,mas (151\,au.) All the disks are small, with only three of them, HD 294260, PDS 113,, and V606 Ori, extending further out than 0.2\arcsec beyond the coronagraph. All the faint disks look smooth and have no visible substructure. RV Ori and PDS 113 have disks with high inclinations ($i>55^{\circ}$.) The disk of PDS 113 is slightly brighter toward the northeast close to the coronagraph than toward the southwest. The disk of RV Ori, though well detected, does not have a well-defined edge and seems to be somewhat more extended to the southwest than to the northeast. The smallest disk, V1650 Ori, is only marginally detected around the coronagraph.   





The disk of HD 294260 is the largest of the faint disks, with a radius of $\sim389$\,mas ($\sim151$\,au.) The disk is smooth and featureless. A brightening can be seen at the eastern edge of the coronagraph. This brightening can be attributed to an imperfectly aligned coronagraph, which allows us to trace slightly further-in disk-structures on this side that are then subject to stronger illumination by the central star. Thus, we might see the inner disk at a slightly smaller radius in the east than in the west. The shape of the disk is seen nearly face-on, at a low inclination, $\sim22^{\circ}$. The disk has an asymmetry that extends a bit further to the northeast than to the southwest and cannot be fully matched with the ellipse expected for a circular disk seen at low inclination. This could thus be an effect of the disk being eccentric. 


A small disk of PDS 110 is visible around the coronagraph in scattered light. An ellipse fitted to the disk extends over the major axis to $187.91$\,mas ($\sim 65$\,au.) The disk seems to be brighter in the southeast than in the northwest, suggesting that the southeastern part is the near side (forward scattering side.) A companion brown dwarf or a planet has been inferred from periodic eclipses where the light intensity dropped by 30 percent (as presented in \citet{2017MNRAS.471..740O} measured in observations from 2008 and 2011.) The unseen companion would be located in a $\sim2$\,au orbit with a mass in the range of 1.8-70$M_{\rm Jup}$ and with a circumsecondary disk of $\sim 0.3$\,au. If additional observations can confirm these eclipses, it would be remarkable, since the planet would be on an orbit that is inclined relative to the measured disk. 

The low-surface-brightness disk, V606 Ori, is viewed face on, and is possibly the largest ($\sim380$\,mas corresponding to $\sim152$\,au) of the faint disks. No features in the disk can be identified at this low surface brightness. It is also not possible to use our automatic routine, described in appendix C, to measure the disk inclination, PA, and radius. We have therefore estimated the radius of the disk and can conclude that it is oriented at a low inclination ($\leq25^{\circ}$.) The low brightness in scattered light could, together with the disk's high millimeter flux, if the disk is not particularly flared, indicate that the bulk of the dust is located in the inner disk so that the dust can effectively shadow the disk.

\subsubsection{Non-detections}

In the remaining 13 targets, we do not detect any scattered light signal that can be interpreted as coming from a disk. The coronagraph covers 92.5\,mas ($\sim30$ to $\sim37$\,au depending on the distance to the targets), which gives an upper limit to the size of any small disks present in these systems. Nine of the 13 targets are systems in which we detect a stellar companion where dynamical interaction could have led to a truncated small disk. In contrast, among the disk detections, only V1650 Ori has a companion. A number of targets, Kiso-A-90460, Brun 252, RY Ori, V499 Ori, and V2149 Ori, have SEDs reminiscent of debris disks.

\subsection{Disk brightness versus stellar and disk properties}
In this section, we compare $\delta_{\rm pol}$ with the SED, age, mass, and 1.3mm flux (F$_{1.3}$) of our target systems to identify correlations between these parameters and the disk brightness (and thus the disk illumination) in scattered light.

\citet{Garufi2018} studied a sample of 58 young planet-forming disks in scattered light and found that systems with large mid- and far-IR excess but low near-IR excess are brighter in scattered light than systems with a strong near-IR excess. This may indicate that disks with large inner cavities are well illuminated, while in disks with high near-IR excess the inner disk material blocks the light to the outer disk region; in other words, these disks are effectively self-shadowed. 
The lack of near-IR excess radiation (with a minimum typically close to 10$\mu$\,m), combined with the presence of strong mid- and far-IR excess, is the typical SED signature of so-called transition disks (\citealt{1989AJ.....97.1451S,2005ApJ...630L.185C, 2007ApJ...670L.135E,2010ApJ...717..441E}.)
To compare the $\delta_{\rm pol}$ with how “transitional” the disks are, we measured the spectral slope between the K band and 5$\mu$m ($\alpha_{K-5}$) and between 5$\mu$m and 22$\mu$m ($\alpha_{5-22}$.) We used \textit{2MASS} photometry for the K band \citep{2003yCat.2246....0C} and \textit{WISE} photometry for 5$\mu$m and 22$\mu$m (WISE band W1 and W4) \citep{2014yCat.2328....0C}. We calculated the slope by using

\begin{equation}
    \alpha_{\lambda}=\frac{\Delta \log(F_{\lambda})}{\Delta \log(\lambda)}.
\end{equation}
 
\begin{figure}[ht]
\centering
\includegraphics[width=8.8cm]{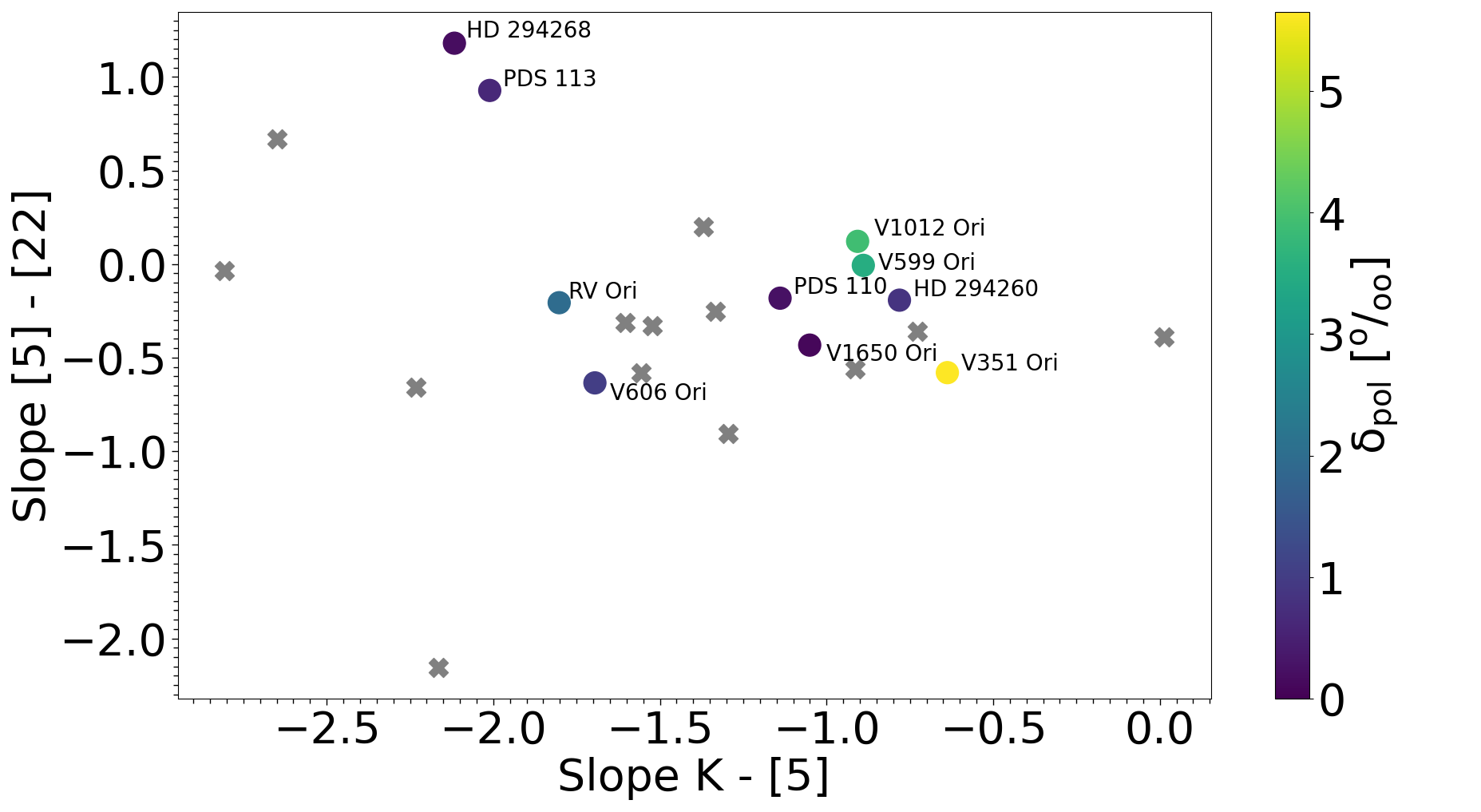}
\caption[]{Comparison between spectral slopes of the SED and the brightness of the disk, $\delta_{\rm pol}$, given by the colorbar. Sources without the detection of a disk in scattered light are marked with a cross.}
\label{fig: poltrans}
\end{figure}

Transitional disks, with a central cavity, have a falling spectral slope between the K band and 5$\mu$m ($\alpha_{K-5}\leq 0$) and a rising slope between 5$\mu$m and 22$\mu$m ($\alpha_{5-22}\geq 0$) \citep{1989AJ.....97.1451S,1990AJ.....99.1187S}. Continuous disks, on the other hand, with small or no cavities, have a constant or falling spectral slope in both ranges. This means that in figure \ref{fig: poltrans} we expect that the more transitional a disk is, the higher and the further to the left it is located in the diagram. 

In the lower left corner we find RV Ori and V606 Ori, which are both low brightness disks. These disks have a steep near-IR slope due to low $5\mu m$ flux. 

The large bright disks in our sample (V\,1012\,Ori, V\,599\,Ori, and V\,351\,Ori) are all located at the right hand end of the diagram, indicating that some near-IR excess, and thus inner disk material close to the star, is present. However, they all show either a positive, flat, or only slightly negative SED slope between 5$\mu$\,m and 22$\mu$\,m, indicating that a large central cavity between the inner disk and the outer (resolved) disk is present. In the case of large disks with moderate to high flaring, slopes could also arise from near-IR self-absorption. 
Only in the case of V\,351\,Ori can such a cavity be confirmed from the scattered light observation (see section 5.2.1.) This large cavity also explains the position of V351\,Ori in figure \ref{fig: poltrans}. The rise in the SED toward the mid- and far-IR (figure \ref{fig: SEDgal}) of V\,351\,Ori starts at longer wavelengths than the spectral slope $\alpha_{5-22}$ measures. 

Of the remaining disks, HD\,294260 is of interest, since it extends well beyond the coronagraph in the observations (0.389\arcsec, 151\,au.) It shows only a slightly negative $\alpha_{K-5}$ and a flat $\alpha_{5-22}$ slope, qualitatively similar to V\,351\,Ori. However, as can be seen in figure~\ref{fig: SEDgal}, it lacks the rise in excess flux toward longer wavelengths, seen in V\,351\,Ori. HD\,294260 indeed shows no clear indication of a significant local minimum in excess flux. In combination with the lack of excess flux toward longer wavelengths, this may indicate either that no large central cavity is present and the outer disk is (partially) self-shadowed by inner disk material, or that the outer disk is dust-depleted. Both scenarios fit well with the fact that we measure a low $\delta_{\rm pol}$ for the system.  

\begin{figure*}
\centering
\includegraphics[width=0.999\textwidth]{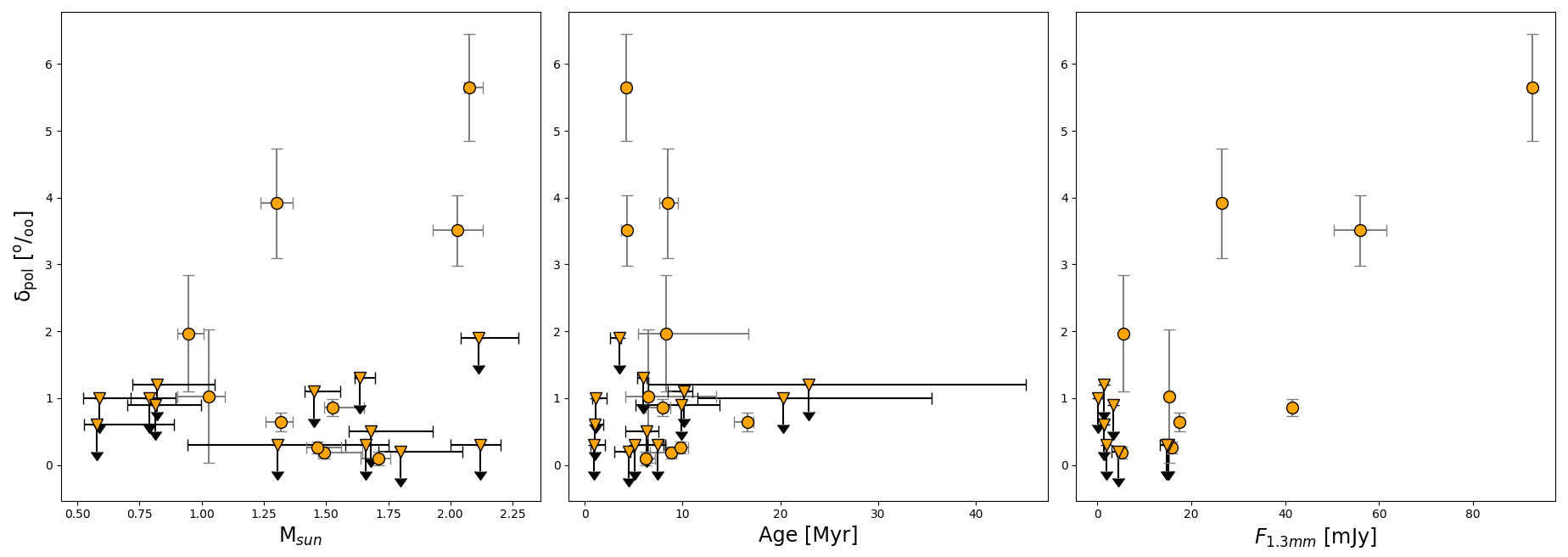}
\caption[]{Polarized-to-stellar light contrast compared to \textbf{a)} Stellar mass, \textbf{b)} Age, and \textbf{c)} 1.3mm Flux.}
\label{fig: threepanel}
\end{figure*}

Comparing $\delta_{\rm pol}$ with the stellar properties, age and mass, we see that the brightest disks happen to be around the most massive and youngest stars in the sample (figure \ref{fig: threepanel}.)  However, the uncertainties in age are large and the sample, out of a PMS perspective, does not contain very young stars, making a correlation with age uncertain. Looking at the brighter portion of the sample, one can see a possible correlation between stellar mass and $\delta_{\rm pol}$. The less bright disk however ($\delta_{pol }\leq 1$) are found around the middle part of the sample mass range. To test this and whether or not there are correlations, we calculated the Kendall $\tau$-coefficient \citep{10.1093/biomet/30.1-2.81}. We find for age $\tau_{K}=0.044$ with a probability of $\sim77\%$ that the $\delta_{\rm pol}$ and age are independent of each other, and for mass $\tau_{K}=0.068$ with a probability of $\sim65\%$ that that the $\delta_{\rm pol}$ and mass are independent of each other. In the sample as a whole, there is therefore no correlation between these quantities.

More interesting is the correlation we see between brightness and the millimeter flux (figure \ref{fig: threepanel} right-most panel.) Bright disks also have a high millimeter flux; that is, they are more massive in dust. To test this correlation we also calculated the Kendall $\tau$-coefficient. We find a correlation between the millimeter flux and $\delta_{\rm pol}$, $\tau_{K}=0.23$, with a probability of $\sim20\%$ that the two quantities are independent of each other. This finding is further discussed in Sect.\,\ref{sec:discussion_faint}.

\section{Discussion} \label{sec:discussion}

\subsection{Disk asymmetries}

We observed 23 systems with IR excess in Orion. In our H-band observations we resolved an extended dusty disk in scattered light in 10 of these 23. Five of these disks have radii over 150\,au; three of these are bright, with V351 Ori being the most extended ($360\pm{20}$\,au) and the brightest. All three of these extended and bright disks in the sample show strong asymmetries in scattered light (see figure \ref{fig:diffdisk}.) These asymmetries could arise from shadows cast by warps in the disk or a misaligned inner disk, as has been suggested for HD\,139614 \citep{2020A&A...635A.121M}, HD\,142527 \citep{2015ApJ...798L..44M}, TW\,Hya \citep{2017ApJ...835..205D}, or HD\,143006 \citep{2018A&A...619A.171B}. 
If the observed asymmetries in our disk sample are indeed caused by shadowing, then the shadows should be azimuthally broad, which would be indicative of a mild warp or small misalignment \citep{2020A&A...635A.121M, 2018A&A...619A.171B}. 
These could be caused by gravitational instability \citep{2016ARA&A..54..271K}, dynamic interactions with stellar companions \citep{2018ApJ...859..118B}, or interaction between an embedded planet and the disk \citep{2018ApJ...869L..47Z}. 

Of the three systems in question, V\,1012\,Ori does not show a visible stellar companion within the field of view of our SPHERE observations (radius of 6", see appendix~\ref{app: mass-limit section} for detection limits.) We detect a faint point source 1.7" to the east of V\,351\,Ori. This object was already detected by \cite{2020AJ....159..252W}, who disregarded it as a companion and classified it as a background field star due to its rather neutral color using Ks and L' photometry. We see no further stellar companion candidates within the SPHERE field of view (see appendix~\ref{app: mass-limit section} for detection limits.) In the case of V599\,Ori we detect a faint companion 2.04\arcsec to the northeast of the primary star. By comparison with VLT/NACO data, taken in 2008, we find that the companion proper motion is consistent with being bound with V599\,Ori. Following examples such as \citet{2018A&A...620A.102D, 2020A&A...635A..73B}, which use the approach of \citet{2014A&A...566A.103L}, we tried to evaluate the probability that the observed companion is a field star. The probability, P, of observing a physically unrelated star in the field at a specific location in the sky is given by

\begin{equation}
P(r,b,m_{\odot},\Delta m_{\rm max})= \pi r^{2} \rho(b,m_{\odot},\Delta m_{\rm max})
\label{eq:prob}
\end{equation}
where r is the separation from the star, b the galactic longitude, $m_{\odot}$ the apparent magnitude of the star in the observed filter, $\Delta m_{\rm max}$ the maximum achieved contrast at the separation, $r$, and $\rho$ the stellar density.  In this case, the possible companion has a separation of 2.05\arcsec and an H-band magnitude of 14.19 mag. To find the stellar density, the amount of stars possibly observed within 1 deg$^{2}$ of V599 Ori, we utilized the \textit{TRILEGAL}\footnote{http://stev.oapd.inaf.it/cgi-bin/trilegal} population synthesis code \citep{2005A&A...436..895G} with default parameters for different parts of the Galaxy and the log-normal initial mass function from \citet{2001ApJ...554.1274C} . \textit{TRILEGAL} estimates a stellar density of 1303 stars per deg$^2$ around V599 Ori with a limiting magnitude of 14.19 mag in the H band. Using Formula \ref{eq:prob} we find a probability of $1.3\times10^{-3}$ that the star by chance could be a field star. Considering the object co-moving with V599 Ori and the low probability of the companion being a field star, we find it very likely that this is a physical companion.
Assuming that the orbit lies in the disk plane, this corresponds to a separation of 977$\pm{68}$\,au. If indeed this is a bound companion, then it may be responsible for the disk asymmetry in V599\,Ori by driving a spiral arm in the disk. However, dedicated hydrodynamic simulations are needed to determine if an orbit configuration consistent with the current projected position brings the companion close enough to the disk to drive a large-scale spiral arm.

The lack of an outer companion in V\,1012\,Ori and V\,351\,Ori suggests that disk gravitational instability or embedded planets may be responsible for the observed asymmetries. If the $M_{\rm disk}/M_*$ are larger than $\sim 10^{-2}$ then gravitational instability could be the reason \citep{2016ARA&A..54..271K}. We therefore estimated the dust disk mass, assuming that there are optically thin conditions at $\lambda=1.3$\,mm ($\nu=230$\,Ghz) and that the dust is isothermal, with
\begin{equation}
M_{\rm dust}=\frac{F_{\nu}D^{2}}{\kappa_{\nu}B_{\nu}(T_{\rm dust})}.
\end{equation}
D is the distance and $B_{\nu}(T_{\rm dust})$ is the Planck function at the dust temperature. We used the dust temperature, $T_{\rm d}=20K$, and $\kappa_{\nu}=2.3\,\mbox{cm}^{2}\,\mbox{g}^{-1}$ for the absorption coefficient. To obtain the total disk mass we then assumed a gas-to-dust ratio of 100. The $M_{\rm disk}/M_*$ then comes out as 0.039 and 0.024 for V351 Ori and V1012 Ori, respectively. Not taking the temperature effects into account, this crude approximation shows that gravitational instability could be a plausible explanation.\\ 
Asymmetry could also arise from the presence of an embedded planet that warps or misaligns the disk \citep{2013A&A...555A.124B, 2013MNRAS.431.1320X, 2018MNRAS.481...20N}, subsequently casting a shadow over the disk. Substellar mass bodies and planets, embedded in the disk, have also been shown to drive vortexes and density waves that induce spiral arms in the disk \citep{2002MNRAS.330..950O, 2012ApJ...748L..22M, 2015ApJ...813...88Z}. Embedded planets open gaps in the disk, trapping dust grains in the pressure bump at the outer edge of the gap \citep{2004A&A...425L...9P, 2014ApJ...785..122Z, 2016A&A...585A..35P}. These processes slow down dust migration and effectively increase the lifetime of the outer disk, resulting in the disk being large and extended for a longer period. Embedded planets in the disks of V351\,Ori, V599\,Ori and V1012\,Ori could thus explain not only the asymmetries observed in scattered light, but also the exceptional radial extent of the disks in these systems.

\subsection{Faint disks and non-detections} \label{sec:discussion_faint}
Faint disks do not show any obvious asymmetries. This is possibly due to our limited sensitivity that does not enable us to detect low-contrast features. The observed correlation between the amount of scattered light and the dust mass probed by the millimeter flux (see Fig.\,\ref{fig: threepanel}) may in principle suggest that all massive disks are properly illuminated. This is however not always seen in a larger, diversified sample \citep[see e.g.,][]{2022A&A...658A.137G} because of the existence of several extended, self-shadowed disks such as those of HD163296 or HK Lup \citep{Garufi2014, 2020A&A...633A..82G}. The absence of such objects in our Orion sample is at the origin of the observed trend and could indicate a more advanced evolutionary stage for the Orion sources, where any large disk that is possibly capable of creating large cavities has done it. 

At the lower end of the dust mass distribution, the absence of any bright disk is partly due to the finite resolution of our observations. In fact, a bright disk of less than 10 mJy may not be resolved by our observations (or may be hidden by the physical coronagraph), and would therefore result in a non-detection. Furthermore, the Orion region is further away than the other regions that are probed by high-contrast imaging (such as Chamaeleon or Taurus), and the fraction of disks that cannot be resolved with 8-m telescopes is larger.


Inspecting the SEDs (see figure \ref{fig: SEDgal}), we can see that many disks have the characteristic shape of the presence of a central cavity. But we only directly image one such cavity in the observations (in the exceptional disk V351 Ori, figure \ref{fig:diffdisk}) and we see substructure in just two of the ten disks. 
Here we are limited by both the resolution (12.25\,mas per pixel) of the telescope and the size of the coronagraph (92.5\,mas.) Cavities that are readily resolved in nearby star-forming regions (in \citet{2016A&A...588A...8G} for example 10 au, HD100456) would be impossible to detect through imaging at a four times greater distance, where the coronagraph size corresponds to 30-37 au.
Because of this, we start to lose a direct connection between SED type and an imaged disk gap. Interferometric observations, with instruments such as MATISSE or GRAVITY, could in the future confirm the presence of central cavities in these targets.

We also see a strong anticorrelation between disk detections and the presence of companions. In nine of the 23 targets, we detect close (candidate) companions. Five of these are first detected in the present study. Out of the nine systems with companions, only two systems also have disks visible in scattered light (V1650 Ori with a relatively small disk and V599 Ori with a large separation to the companion. That disks are small and faint in close binary systems is expected from theoretical predictions of the effects of binaries on disk radii, radial drift, and dispersion \citep{1993prpl.conf..749L,1994ApJ...421..651A, 2021MNRAS.501.2305Z} and earlier observational evidence at millimeter wavelengths of close binaries ($\leq 140$\,au) \citep{2012ApJ...751..115H, 2019ApJ...872..158A}. Only two of the binaries have a separation of several $\sim100$\,au, Kiso A-0904 60 and V599 Ori, where the latter as mentioned before harbours a well-defined extended disk. 

In the system V2149 Ori, we measure an increase in separation between the components compared to measurements from \citet{2006A&A...458..461K}. This increase in separation corresponds to a projected velocity of $\sim 3.6\,\mbox{km\,s}^{-1}$. The velocity is such that, given the uncertainties in orbital inclination and eccentricity, one cannot exclude a bound orbital configuration. Further observations of this system in years to come would provide better evidence and may determine if the companion is indeed in a bound orbit.

\subsection{UV-radiation}
\begin{figure}
\centering
\includegraphics[width=8.8cm]{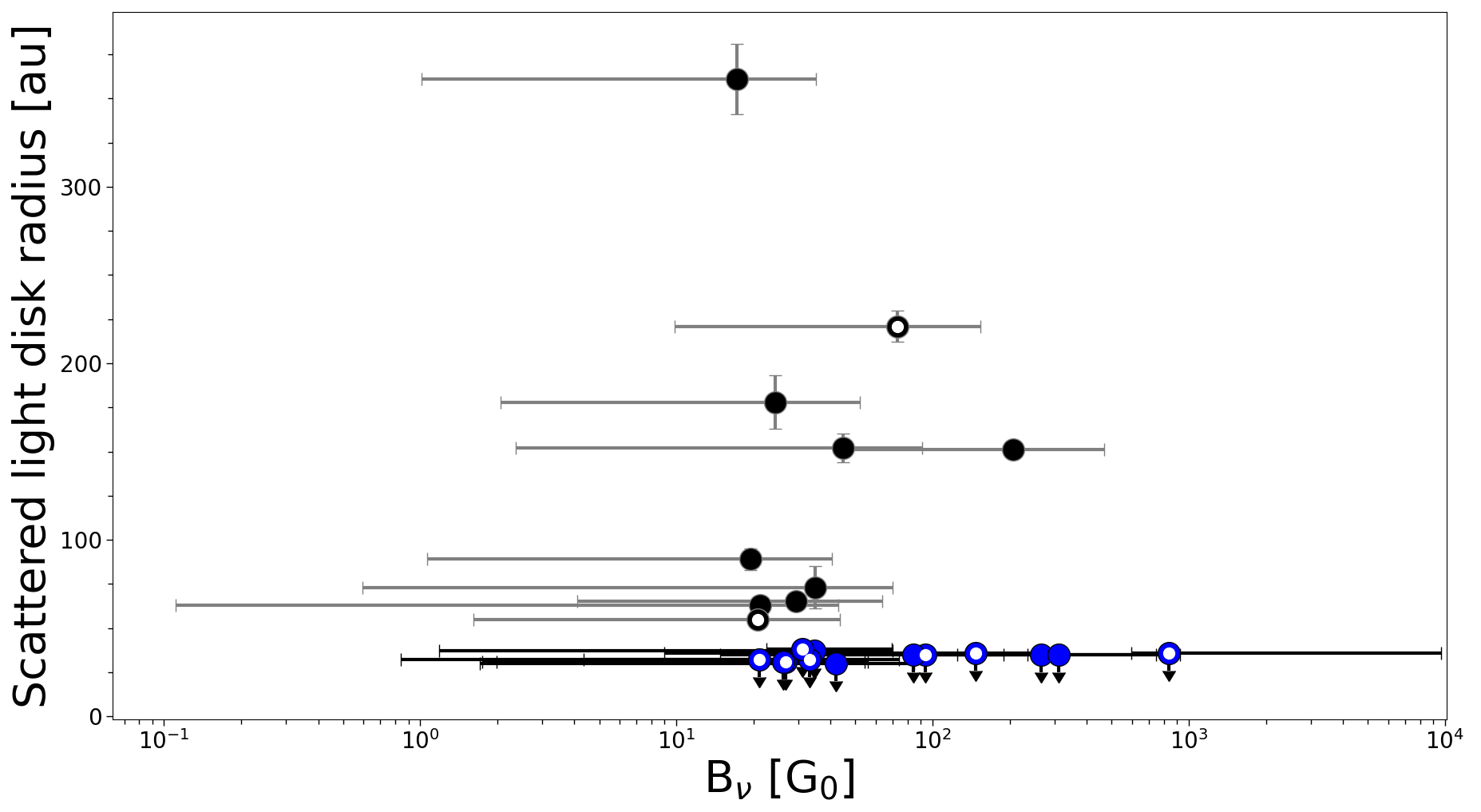}
\caption[]{Graph showing the disk radius of each source in the FUV field of the three O stars, $\theta_{1}$\,Ori, $\sigma$\,Ori, and NU Ori. Local extinction has not been taken into account. Systems with detected disks are black and systems without a detected disk are blue. Upper limits for non-detections were derived using the angular size of the coronagraph. Binary systems are indicated with a white marker. No disks are detected at FUV fields higher than $\sim 300 G_{0}$. }
\label{fig: UVinfluence}
\end{figure}

Finally, we investigated if there is a correlation between the local far-ultraviolet (FUV) field and the measured disk radius in scattered light. Using the distances from \citet{2021AJ....161..147B}, we calculated the geometric distances of each sample star from the three O stars, $\theta_{1}$\,Ori, $\sigma$\,Ori, and NU Ori, which can be assumed to be the main contributors to the FUV field in the region. Using a standard black body and stellar models for massive stars \citep{2005A&A...436.1049M}, we integrated between 91-200 nm to obtain an estimate of the $L_{\rm FUV}$ for each of the O stars. We then calculated the total FUV field for each location in our sample by adding the contributions from all O stars. We did not take local extinction into account when doing this, since we were only interested in seeing if there is a general influence from the FUV field. As we see in figure \ref{fig: UVinfluence}, the detected disks are found at FUV fields lower than $\sim 300 G_{0}$, with the most extended disk (around V351\,Ori) being exposed to the lowest FUV field. Whereas \citet{2022A&A...661A..53V} finds a weak trend between the disk radius and UV field strength of young objects in L1641 and L1647, we see no clear trend between the scattered light disk radius and external FUV field. In this case, the disks resemble the one close by LM-SFR (for example Taurus and Lupus) where in the absence of massive O and B stars it is unlikely that the FUV field is very strong. Even though no clear correlation is visible, we note that the largest disks in the sample can be found where the UV field strength is low. This may explain the larger fraction of compact disks we observe in contrast to the samples of disks in scattered light located in LM-SFR (see \citet{2018ApJ...863...44A} for example).  


\section{Summary} \label{sec:conclusions}
We have presented the first comprehensive study of bright disks in Orion with extreme adaptive optics high-contrast imaging. 23 systems were observed with \textit{VLT/SPHERE} in the H band (1.6 $\mathrm{\mu m}$) in polarized
light. The observed disks (with the exeptions of V351 Ori, V599 Ori, and V1012 Ori) are generally compact in comparison to the disks observed in LM-SFR. This is shown both in the scattered light radius but also in the correlation between the amount of scattered light and the dust mass. This may be explained by a more advanced evolutionary state of large disks but also, even though no strong environmental effects are seen in this sample, by a stronger UV field. The strong anticorrelation between disk detection and the presence of a companion shows that the higher frequency of companions in HM-SFR has a significant impact on the frequency of disks, since it could speed up disk clearing and prevent the formation of large disks.  In the following, we summarize the main results of this survey:

\begin{enumerate}
    \item We detect 10 protoplanetary disks, out of which three are bright and extended (V351\,Ori, V599\,Ori, and V1012\,Ori.)
    \item The bright disks are asymmetric and two of these (V351 Ori and V599 Ori) show substructure.
    \item We find a correlation ($\tau_{K}=0.23$) between the 1.3\,mm flux and the scattered light brightness ($\delta_{\rm pol}$) of the detected disks. This correlation is due to the absence of large, self-shadowed disks.
    \item We detect nine systems harbouring close stellar companions, of which five were previously unknown.
    \item We additionally detect a possible substellar companion to RY\,Ori located at 137\,au from the primary star.
    \item We find a strong correlation between multiplicity and the absence of a disk detection. We detect only two disks among the nine detected multiple systems.
    \item We lose the direct connection between inner SED-inferred disk cavities and bright scattered light disks, likely as a consequence of the limited angular resolution at the distance of the Orion SFR.
    \item We see no direct environmental effects on disk radii as a result of a more intense external FUV field. In this sense, the sampled disks resemble those in LM-SFR. 
    
\end{enumerate}

Our study demonstrates the value of high-resolution near-IR observations in regions up to 400\,pc distance with current instrumentation. Once 30\,m class telescopes such as the ESO-ELT become available it would be most interesting to follow up on the many angular compact disks discovered in our sample to study the presence of substructures and compare these to the exceptionally extended disks in the same region.
  
\begin{acknowledgements}
This publication is part of the project The precursors of evolved protoplanetary disks around Herbig Ae/Be stars (with project number 023.012.014) of the research programme Promotiebeurs voor Leraren which is financed by the Dutch Research Council (NWO.)
\\
This work has made use of data from the European Space Agency (ESA) mission
{\it Gaia} (\url{https://www.cosmos.esa.int/gaia}), processed by the {\it Gaia}
Data Processing and Analysis Consortium (DPAC,
\url{https://www.cosmos.esa.int/web/gaia/dpac/consortium}.) Funding for the DPAC
has been provided by national institutions, in particular the institutions
participating in the {\it Gaia} Multilateral Agreement.
\\
This paper makes use of the following ALMA data: ADS/JAO.ALMA\#2021.1.01705.S. ALMA is a partnership of ESO (representing its member states), NSF (USA) and NINS (Japan), together with NRC (Canada), MOST and ASIAA (Taiwan), and KASI (Republic of Korea), in cooperation with the Republic of Chile. The Joint ALMA Observatory is operated by ESO, AUI/NRAO and NAOJ.
\\
\'A. Ribas has been supported by the UK Science and Technology research Council (STFC) via the consolidated grant ST/W000997/1 and by the European Union’s Horizon 2020 research and innovation programme under the Marie Sklodowska-Curie grant agreement No. 823823 (RISE DUSTBUSTERS project.)
\\
T. Birnstiel acknowledges funding from the European Research Council (ERC) under the European Union’s Horizon 2020 research and innovation programme under grant agreement No 714769 and funding by the Deutsche Forschungsgemeinschaft (DFG, German Research Foundation) under grants 361140270, 325594231, and Germany's Excellence Strategy - EXC-2094 - 390783311.
Support for J. Huang was provided by NASA through the NASA Hubble Fellowship grant \#HST-HF2-51460.001-A awarded by the Space Telescope Science Institute, which is operated by the Association of Universities for Research in Astronomy, Inc., for NASA, under contract NAS5-26555.
\\
C. Manara. Funded by the European Union (ERC, WANDA, 101039452.) Views and opinions expressed are however those of the author(s) only and do not necessarily reflect those of the European Union or the European Research Council Executive Agency. Neither the European Union nor the granting authority can be held responsible for them.
\\
Ch. Rab is greatful for the support from the Max Planck Society.
\\
A. Zurlo acknowledges support from ANID -- Millennium Science Initiative Program -- Center Code NCN2021\_080
\end{acknowledgements}
%
%

\bibliographystyle{aa} 
\bibliography{references} 

\begin{appendix}

\section{Measuring inclination, PA, and radius} \label{sec:disk_measurement}

For all disk detections within our sample (with the exception of the complex substructured disk surrounding V351\,Ori) we used ellipse fitting to the outer disk edge to determine the basic disk geometry. The fitting procedure is described in detail in the companion paper to this study on the Cha\,I star-forming region (Ginski et al., submitted.) Here, we give a brief summary. 

To determine the radial location of the disk edge we extracted radial profiles using a sliding aperture with a diameter of 4 pixels; that is, roughly corresponding to the size of the resolution element in the H band. For each radial profile, we determined the point at which the disk flux drops below $3\sigma$ of the noise background of the image and used this as the edge position. In order to estimate the uncertainty of thsi edge location, we used a sub-array centered around the disk edge and fit the linear slope of the profile. We then scaled the uncertainties with the inverse of the slope. The general idea is that the edge location will be very clear for a sharply truncated disk with a steep slope in signal across the edge, but be more diffuse for an extended disk that slowly fades below the background noise due to, for example, the drop in illumination at larger distances from the star.

Once the data points were extracted, we used the least squares algorithm developed by \cite{oy1998NUMERICALLYSD} to obtain the ellipse solution that best fit the data points. To explore the uncertainties of this solution we repeated this process $10^5$ times, each time drawing the data points individually from a normal distribution centered on their original position and with a width based on their uncertainties. 

\begin{figure*}
\centering
\subfloat[HD\,294260]{
\includegraphics[scale=0.3]{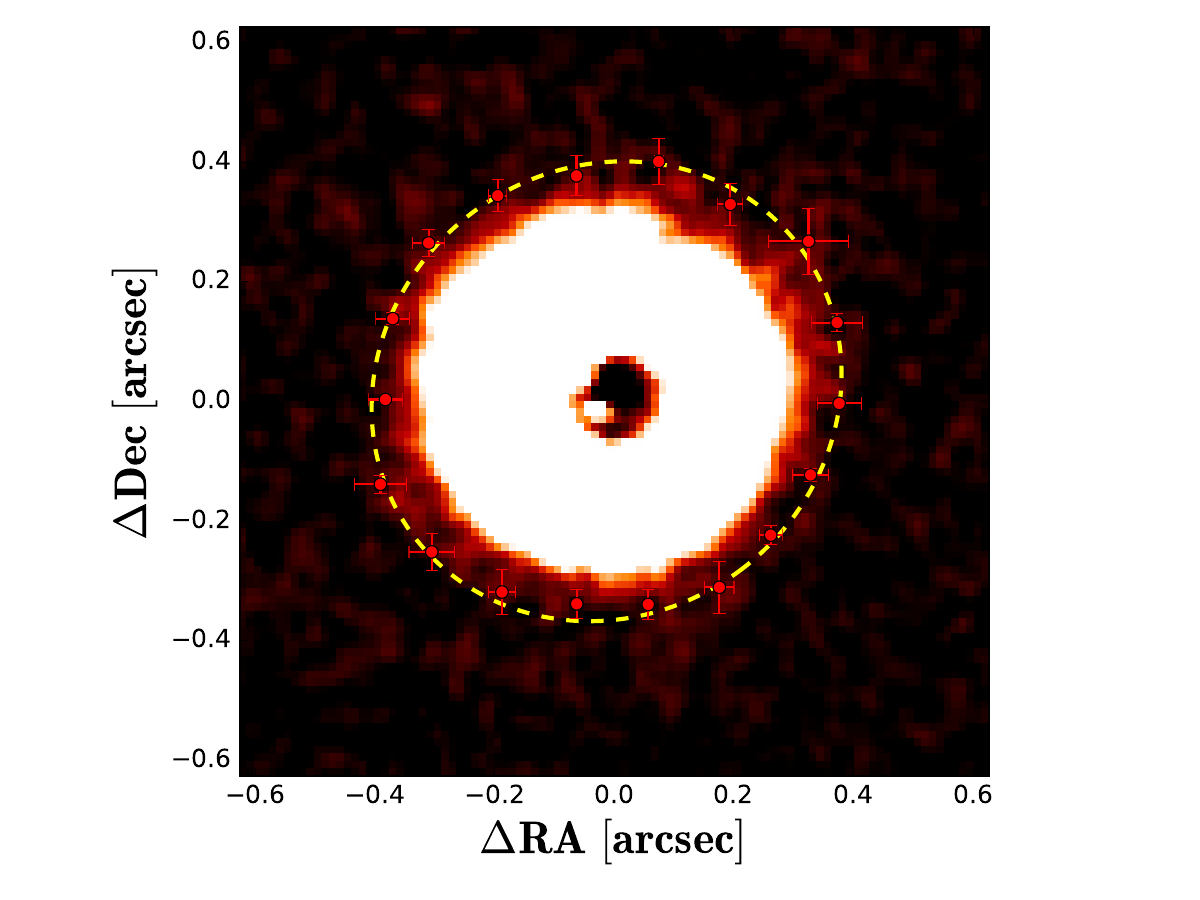}
\label{HD294260-ellipse}
}
\subfloat[HD\,294268]{
\includegraphics[scale=0.3]{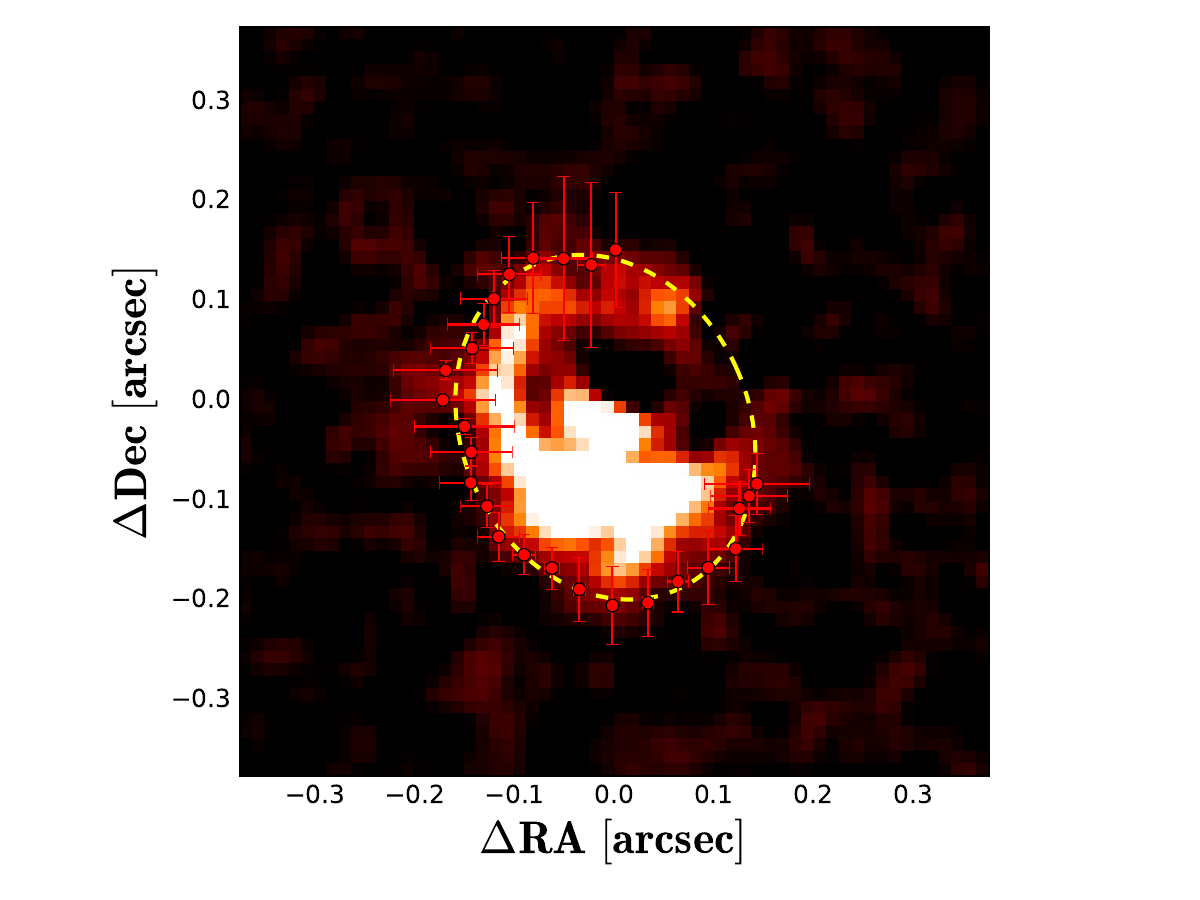}
\label{HD294268-ellipse}
}
\subfloat[PDS\,110]{
\includegraphics[scale=0.3]{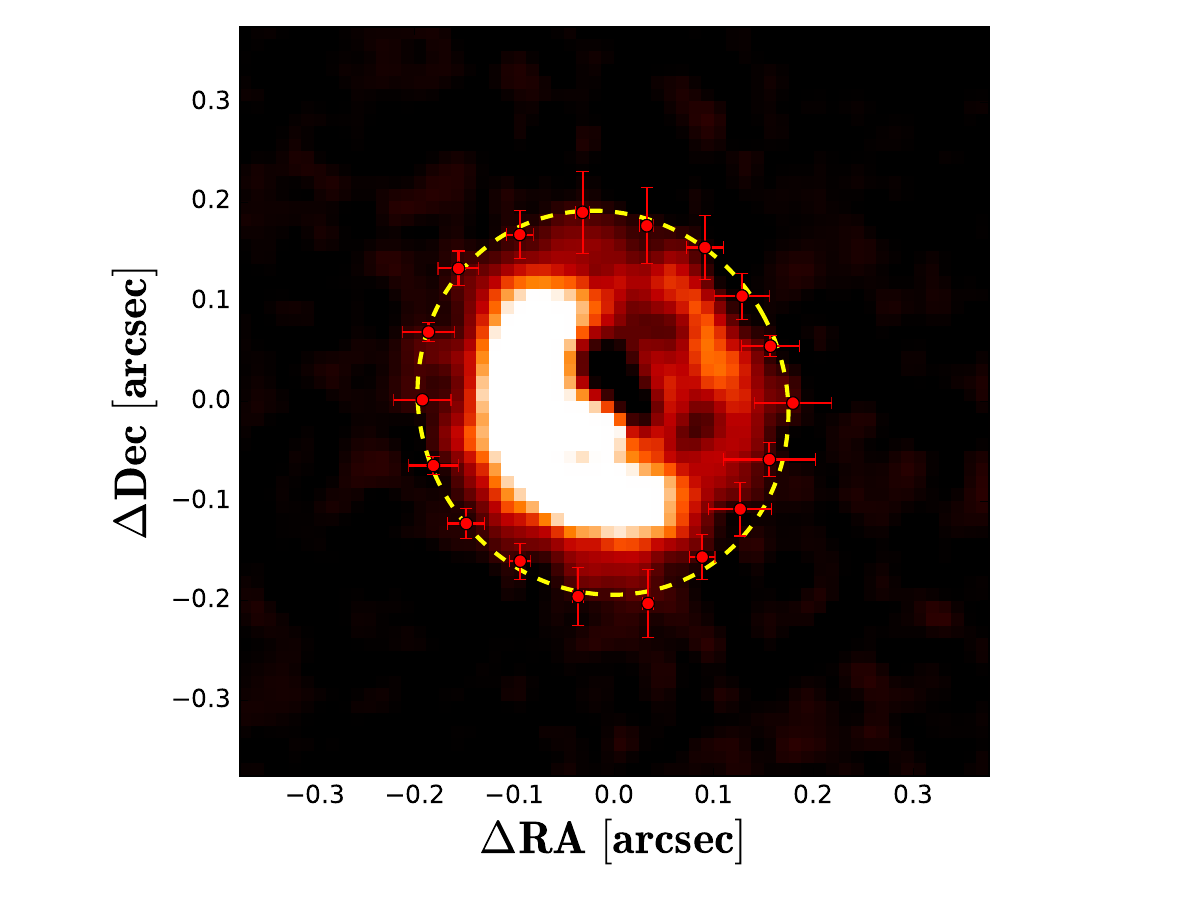}
\label{PDS110-ellipse}
}

\subfloat[PDS\,113]{
\includegraphics[scale=0.3]{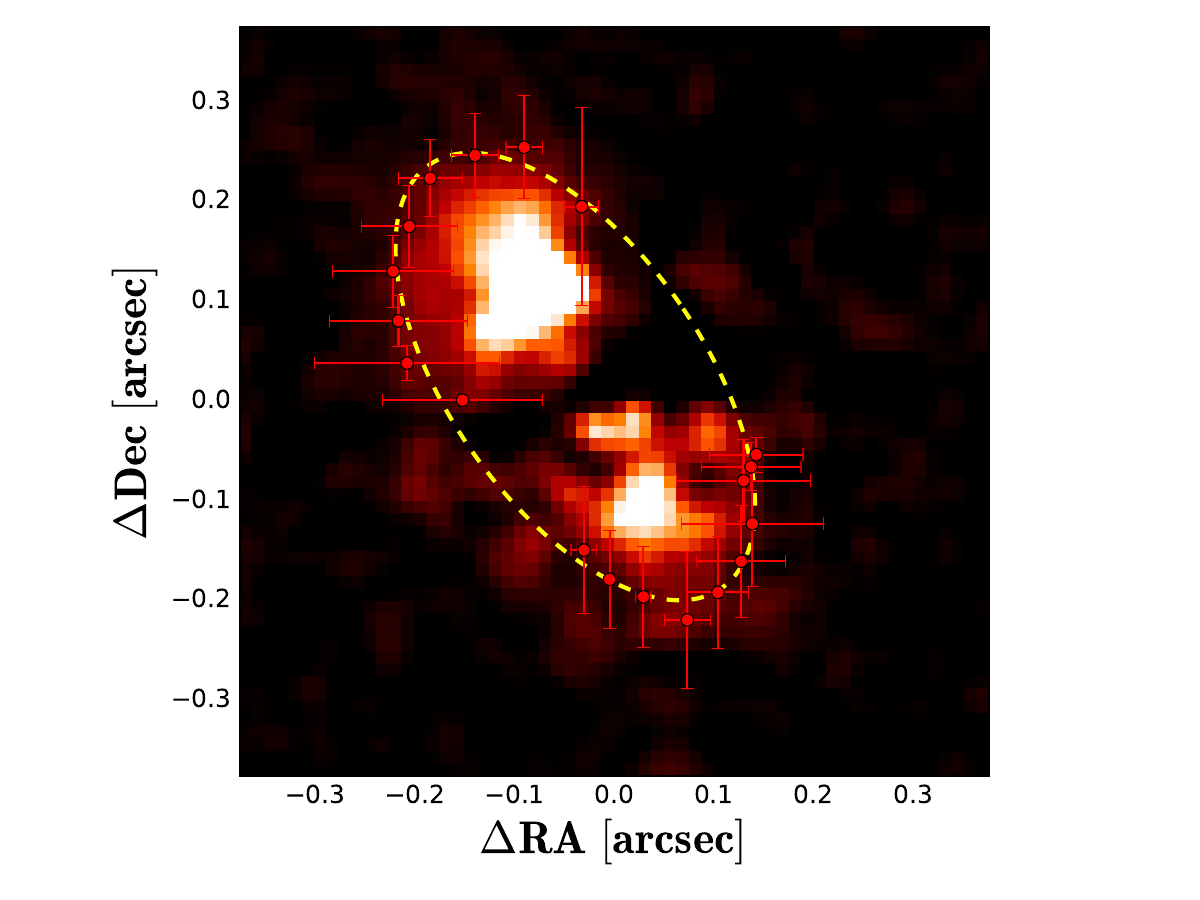}
\label{PDS113-ellipse}
}
\subfloat[RV\,Ori]{
\includegraphics[scale=0.3]{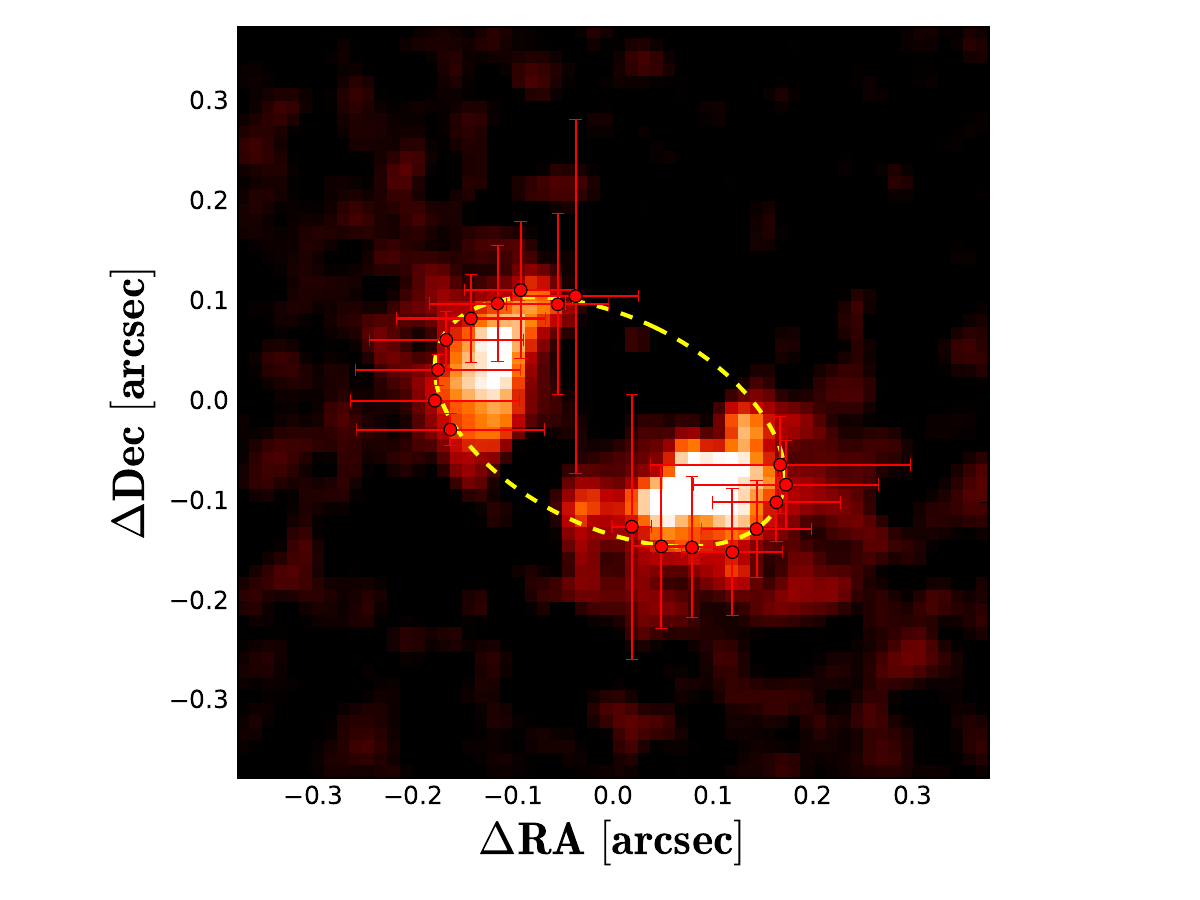}
\label{RVOri-ellipse}
}
\subfloat[V599\,Ori]{
\includegraphics[scale=0.3]{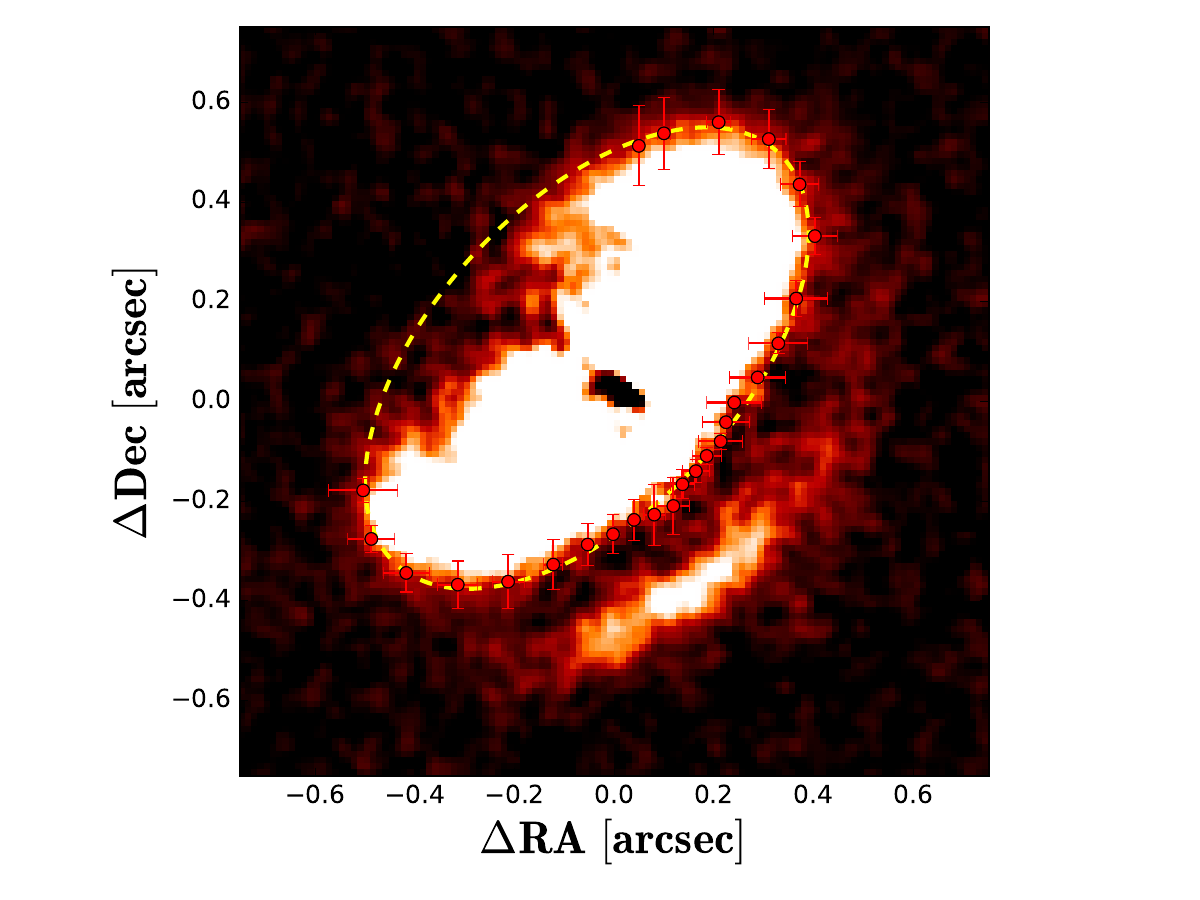}
\label{V599Ori-ellipse}
}

\subfloat[V1012\,Ori]{
\includegraphics[scale=0.3]{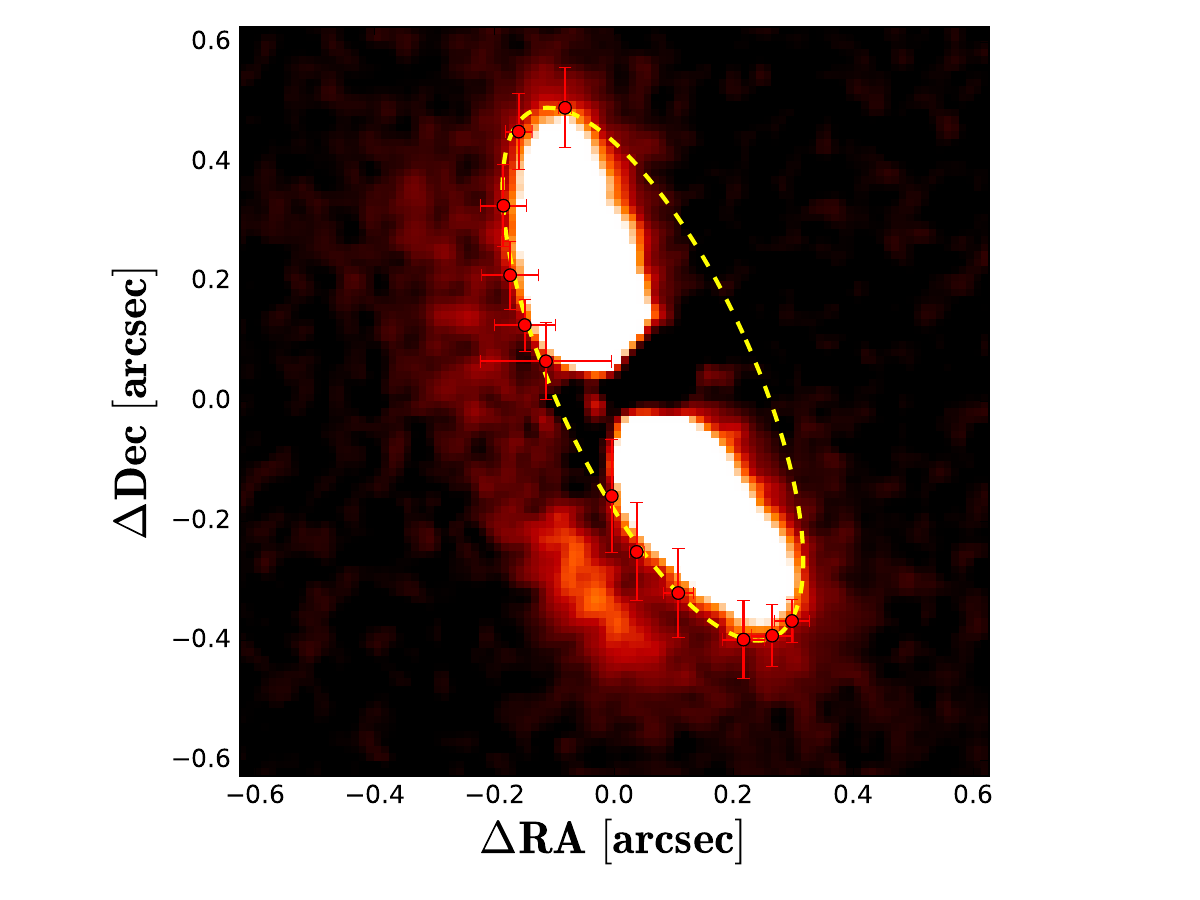}
\label{V1012Ori-ellipse}
}
\subfloat[V1650\,Ori]{
\includegraphics[scale=0.3]{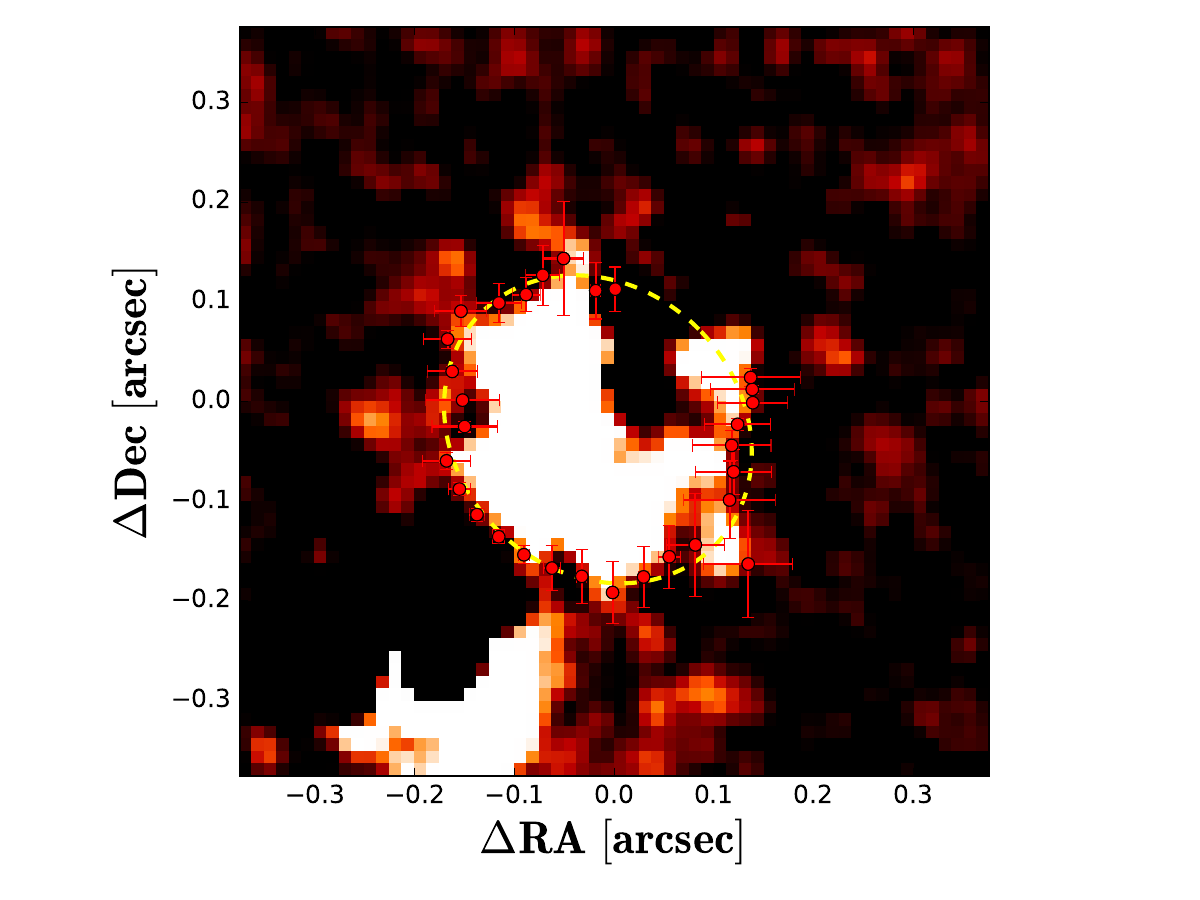}
\label{V1650Ori-ellipse}
}

\caption[]{Ellipse fit to the outer disk edge in eight systems in this study. Disk images are shown on a saturated color map to highlight the edge, i.e., the region where the disk signal drops below 3$\sigma$ above the sky background. Dashed yellow lines show the final fitted ellipse. }
\label{fig: ellipse-fit}
\end{figure*}

In figure~\ref{fig: ellipse-fit} we show the disk images on a saturated color map to highlight the location of the disk edge. We overplot the extracted locations of the disk edge with their associated uncertainties as well as the best fitting ellipse solution. We note that we excluded regions from the extraction and fit in which the disk signal is very weak, either because it is obscured by the coronagraphic mask (e.g., PDS\,113 and RV\,Ori), or because of the intrinsic scattering phase function or shadowing (e.g., V599\,Ori and V1012\,Ori.)

\FloatBarrier

\section{Spectral energy distributions of the targets}

\begin{figure*}
\centering
\includegraphics[width=0.89\textwidth]{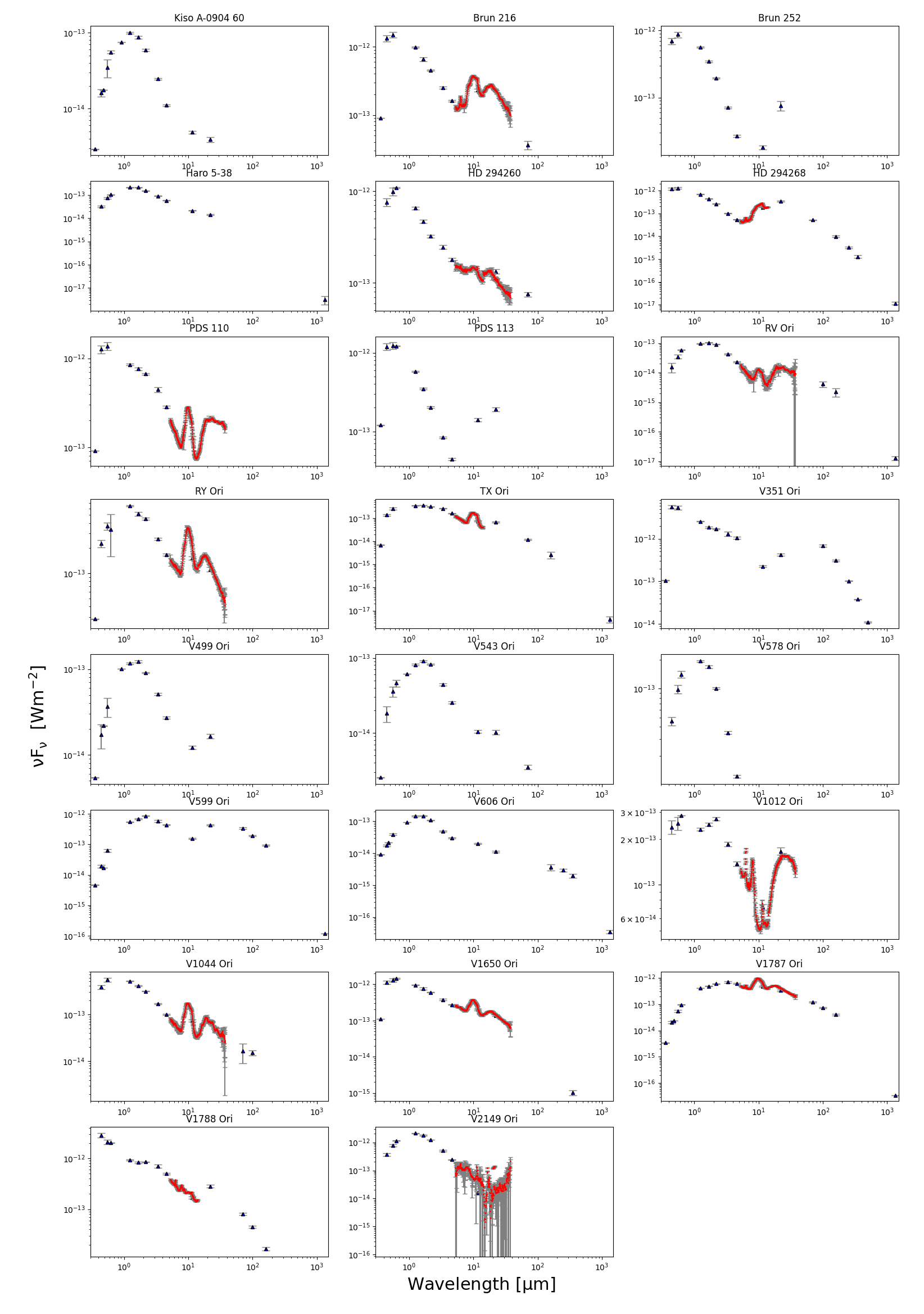}
\caption[]{Spectral energy distributions of the sources in the sample. The photometry and AOR-keys for the CASSIS Spitzer database are available in table \ref{tab:phot}.}
\label{fig: SEDgal}
\end{figure*}

Here we present the SEDs of all the targets in figure \ref{fig: SEDgal}. We list references to all photometry in table \ref{tab:stars} per target.

\begin{table*}
\caption{References to photometry \& spectra.}
\label{tab:phot}\tiny      
\centering             
\begin{tabular}{lllllllll}
Name            & Optical              & 2MASS  & WISE   & Herchel PSC & AOR-Key    & $F_{1.3mm}$ [mJy]         & Ref. 1.3mm                & Ref. photometry             \\
\hline
Kiso A-0904 06  & II/336, V/139, I/327 & II/246 & II/328 &             &            &$\leq0.25$      & This work                 &  1,2,3,5,6             \\ 
Brun 216        & I/322A, V/139        & II/246 & II/328 & VIII/106    &  18832640  &$4.44\pm{1.47}$            & This work                 &  2,4,5,6,7            \\
Brun 252        & I/322A               & II/246 & II/328 &             &            &                           &                           &  4,5,6\\
Haro 5-38       & I/322A, I/327        & II/246 & II/328 &             &            &$1.4\pm{0.15}$             & Ansdell et al. 2017       &  3,4,5,6    \\
HD 294260       & I/322A, I/327        & II/246 & II/328 & VIII/106    &  21878528  &$41.51\pm{0.12}$           & This work                 &  3,4,5,6,7             \\
HD 294268       & I/322A               & II/246 & II/328 & VIII/106    &  18147328  &$5.16\pm{0.13}$            & Ansdell et al. 2017       &  4,5,6,7  \\
PDS 110         & I/322A, V/139        & II/246 & II/328 &             &  21889024  &$15.94\pm{0.08}$           & This work                 &  2,4,5,6            \\
PDS 113         & I/322, V/139, I/327  & II/246 & II/328 &             &            &$17.54\pm{0.08}$           & This work                 &  2,3,4,5,6              \\
RV Ori          & II/336, I/327        & II/246 & II/328 & VIII/106    &  26309888  &$5.66\pm{0.15}$            & Ansdell et al. 2017       &  1,4,5,6,7\\
RY Ori          & II/336, V/139, I/327 & II/246 & II/328 &             &  21871616  &                           &                           &  1,2,3,5,6    \\
TX Ori          & I/322A, V/139        & II/246 & II/328 & VIII/106    &  18146560  &$1.9\pm{0.13}$             & Ansdell et al. 2017       &  2,4,5,6,7  \\
V351 Ori        & I/322A, V/139        & II/246 & II/328 & VIII/106    &            &$92.59\pm{0.26}$           & This work                 &  2,4,5,6,7\\
V499 Ori        & II/336, V/139        & II/246 & II/328 &             &            &$3.33\pm{0.09}$            & This work                 &  1,2,5,6   \\
V543 Ori        & II/336, V/139, I/327 & II/246 & II/328 & VIII/106    &            &$1.34\pm{0.08}$            & This work                 &  1,2,3,5,6,7   \\
V578 Ori        & I/322A, I/327        & II/246 & II/328 &             &            &                           &                           &  3,4,5,6  \\
V599 Ori        & I/322A, V/139        & II/246 & II/328 & VIII/106    &            &$55.9\pm{5.6}$             & Ansdell et al. 2017       &  2,4,5,6,7   \\
V606 Ori        & I/322A, V/139        & II/246 & II/328 & VIII/106    &            &$15.38\pm{0.25}$           & Ansdell et al. 2017       &  2,4,5,6,7   \\
V1012 Ori       & I/322A, I/327        & II/246 & II/328 &             &  25731584  &$26.48\pm{0.10}$           & This work                 &  3,4,5,6\\
V1044 Ori       & I/322                & II/246 & II/328 & VIII/106    &  21872640  &                           &                           &  4,5,6,7\\
V1650 Ori       & I/322, V/139, I/327  & II/246 & II/328 & VIII/106    &  21870848  &                           &                           &  2,3,4,5,6,7\\
V1787 Ori       & I/322, V/139, I/327  & II/246 & II/328 & VIII/106    &  18834176  &$14.8\pm{1.48}$            & van Terwisga et al. 2022  &  2,3,4,5,6,7\\
V1788 Ori       & I/322A, I/327        & II/246 & II/328 & VIII/106    &  11002112  &$15.13\pm{0.09}$           & This work                 &  3,4,5,6,7\\
V2149 Ori       & I/322A, I/327        & II/246 & II/328 &             &  18802176  &                           &                           &  3,4,5,6\\
\hline
\end{tabular}\\
\smallskip
\normalsize
\raggedright
Table containing the catalogs from Vizier with the photometry used to create the SEDs as well as the B- and V-band fluxes used to derive the luminosity of the sources. 1.3 mm fluxes are referenced directly in the table and  references to all other catalogs can be found below the table. \textbf{References:} 1)II/336 \citep{2015AAS...22533616H}, 2)V/139 \citep{2012ApJS..203...21A}, 3)I/327 (Carlsberg Meridian Catalog Number 15 2011), 4)I/322A \citep{2013AJ....145...44Z}, 5)II/246 \citep{2003yCat.2246....0C}, 6)II/328 \citep{2014yCat.2328....0C}, 7)VIII/106 \citep{2020yCat.8106....0H} 
\end{table*}  

\FloatBarrier
\section{Mass detection limits on V351\,Ori and V1012\,Ori from the SPHERE imaging data}

\label{app: mass-limit section}

\begin{figure}
\centering
\subfloat[V351\,Ori]{
\includegraphics[scale=0.55]{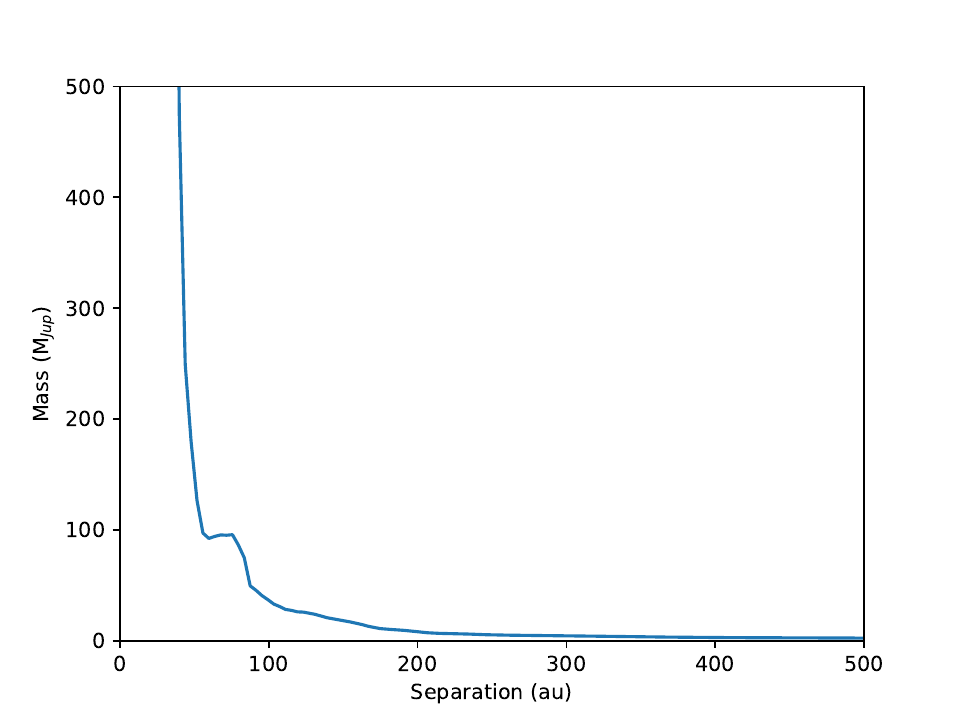}
\label{V351ori-mass-limit}
}

\subfloat[V1012\,Ori]{
\includegraphics[scale=0.55]{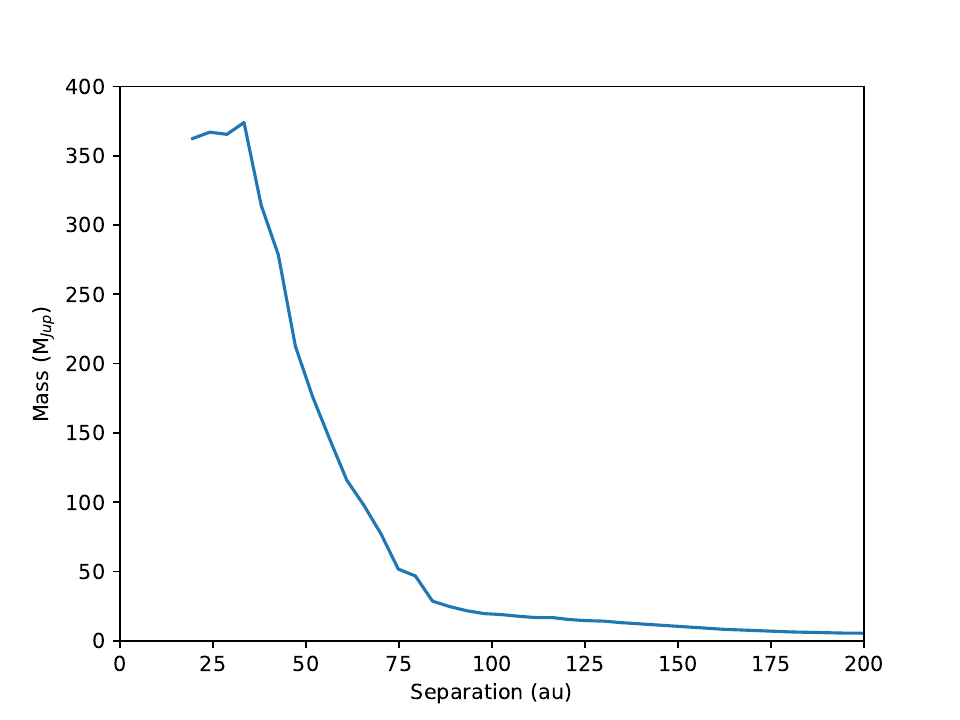}
\label{V1012ori-mass-limit}
}

\caption[]{Mass limits computed from the SPHERE H-band data using angular differential imaging to detect total intensity sources around the two exceptionally extended disks in the V351\,Ori and V1012\,Ori systems. }
\label{app:mass-limits}
\end{figure}

Using the SPHERE imaging data, we produced mass detection limits for the V351\,Ori and V1012\,Ori systems, analogous to what was done in \cite{2022NatAs...6..751C}. We show the results in figure~\ref{app:mass-limits}.

\end{appendix}

\end{document}